\documentclass[pra,aps,epsfig,psfig,multicols,showpacs,tightenlines,onecolumn]{revtex4}
\usepackage{graphics,bm}
\usepackage{graphicx}
\usepackage{amsmath, amssymb, graphics}
\allowdisplaybreaks

\newcommand{\beq}{\begin{equation}}
\newcommand{\eeq}{\end{equation}}
\newcommand{\bqa}{\begin{eqnarray}}
\newcommand{\eqa}{\end{eqnarray}}
\def\gsim{\mathrel {\vcenter {\baselineskip 0pt \kern 0pt
\hbox{$>$} \kern 0pt \hbox{$\sim$} }}}
\def\lsim{\mathrel {\vcenter {\baselineskip 0pt \kern 0pt
\hbox{$<$} \kern 0pt \hbox{$\sim$} }}}

\protect

\begin{document}

 \title{Superposition principle and composite solutions to coupled nonlinear Schr\"odinger equations}

\date{}

\author{L. Al Sakkaf and U. Al Khawaja}

\address{
\it Department of Physics, United Arab Emirates University, P.O.
Box 15551, Al-Ain, United Arab Emirates}

%\newpage

\begin{abstract} \noindent We show that the superposition principle applies to coupled nonlinear Schr\"odinger equations with cubic nonlinearity where exact solutions may be obtained as a linear combination of other exact solutions. This is possible due to the cancellation of cross terms in the nonlinear coupling. First, we show that a {\it composite} solution which is a linear combination of the two components of a {\it seed} solution is another solution to the same coupled nonlinear Schr\"odinger equation. Then, we show that a linear combination of two composite solutions is also a solution to the same equation. With emphasis on the case of Manakov system of two-coupled nonlinear Schr\"odinger equations, the superposition is shown to be equivalent to a rotation operator in a two-dimensional function space with components of the seed solution being its coordinates. Repeated application of the rotation operator, starting with a specific seed solution, generates a series of composite solutions which may be represented by a generalized solution that defines a family of composite solutions.  Applying the rotation operator to almost all known exact seed solutions of the Manakov system, we obtain for each seed solution the corresponding family of composite solutions. Composite solutions turn out, in general, to possess interesting features that do not exist in the seed solution. Using symmetry reductions, we show that the method applies also to systems of $N$-coupled nonlinear Schr\"odinger equations. Specific examples for the three coupled-nonlinear Schr\"odinger equation are given.       
 \end{abstract}

\maketitle

%\begin{center} %\textbf{AMS} Codes: 78A60, 37K10; 35Q51, 35Q55\\
%\textbf{PACS} Codes: 05.30.Jp; 05.45.Yv; 02.30.Ik. %\\ \textbf{OCIS}
%Codes: 060.2310; 060.4510; 060.5530; 190.3270; 190.4370\\
%\textbf{Key words}: solitary waves; traveling waves; integrability;
%numerics. \end{center}

\section{Introduction} \label{intsec}

Superposition principle applies to linear differential equations giving rise to a generalised solution in the form of a linear combination of the independent solutions. Generally, the superposition principle does not apply to nonlinear differential equations. Nonetheless, it was found to apply for specific cases, such as periodic solutions to the Korteweg-de Vries (KdV) equation, modified Korteweg-de Vries (mKdV), and $\lambda\,\phi^4$ field equation  \cite{kharepre}, Kadomtsev-Petviashvili (KP) equation, the nonlinear Schr\"odinger equation (NLSE), sine-Gordon equation, and the Boussinesq equation \cite{kharejpa}. This was possible due to cyclic identities satisfied by the Jacobi elliptic functions where the nonlinear cross terms reduce to, and combine with, other linear terms  \cite{kharecyclic}. Other examples include time modes of nonlinear systems \cite{time modes}, the Novikov-Veselov equation \cite{nve}, Maxwell-Schr\"odinger equations \cite{mse}, the (2+1)-dimensional KdV Equations \cite{2de}, Caudrey-Dodd-Gibbon-Sawada-Kotera (CDGSK) equation and the (2+1)-dimensional Nizhnik-Novikov-Veselov (NNV) equation \cite{jun}, higher order NLSE \cite{hnlse}, coupled nonlinear Klein-Gordon and Schr\"odinger equations \cite{kgse}, the (2+1)-
dimensional modified Zakharov-Kuznetsov equation and the (3+1)-dimensional Kadomtsev-Petviashvili equation \cite{zke}, the (2+1)-dimensional Zakharov-Kuznetsov (ZK) equation and the Davey-Stewartson (DS) equation \cite{dse}, generalized KdV equation, the Oliver water wave equation, the $k(n; n)$ equation \cite{gkdve}, fifth-order KdV equation \cite{5kdv}, cubic-quintic NLSE \cite{cqe}, and (2+1)-dimensional Boussinesq Equation \cite{be}.

The main property that allows for the application of superposition principle to all of the above-mentioned nonlinear systems is the reduction of the nonlinear cross terms into linear ones which then combine with other linear terms. Hence, the nonlinear equation reduces to a sum of two uncoupled replicates for each of the superimposed solutions individually. This is possible only for certain types of solutions that have this property. In Ref.  \cite{kharecyclic}, such solutions were identified as the Jacobi elliptic functions satisfying cyclic identities. Therefore, all solutions obtained using this kind of linear superposition were periodic. 

Here, we report a different method allowing for linear superposition principle in the $N$-coupled nonlinear Schr\"odinger equation, where the cross terms cancel each other. The possibility of cancelling the cross terms is brought by the structure of the nonlinear term and is independent of the solutions used. Therefore, this kind of superposition principle applies, not only to periodic solutions, but also to localized solutions such as dark and bright solitons which will introduce interesting new features.

Consider the two-coupled NLSE with cubic nonlinearity of the form $|\psi_1|^2+|\psi_2|^2$, where $(\psi_1,\psi_2)^T$ is a known {\it seed} solution to the coupled NLSE. When a linear combination of the two components $(\phi_1,\phi_2)^T=(c_1\,\psi_1+c_2\,\psi_2,c_3\,\psi_1+c_4\,\psi_2)^T$, which we denote here and throughout as a {\it composite solution}, is substituted back into the nonlinear term, it will be possible for certain choices of the constants $c_{1-4}$ to cancel all cross terms and eventually to obtain, out of each part of the coupled NLSE, two uncoupled NLSEs for $\psi_1$ and $\psi_2$. Since $\psi_1$ and $\psi_2$ already assumed to satisfy the coupled NLSE, then $\phi_1$ and $\phi_2$ will also do. Although, we will argue below that this  indeed  shows the applicability of the superposition principle, it will be possible to verify, in a more straightforward manner, that a linear superposition of composite solutions is also a solution, namely $(\Phi_1,\Phi_2)^T=d_1\,(\phi_1,\phi_2)^T + d_2\,(\phi_3,\phi_4)^T$, where $d_1$ and $d_2$ are arbitrary constants, is a solution to the coupled NLSE, if $(\phi_1,\phi_2)^T$ and $(\phi_3,\phi_4)^T$ are two solutions to the same coupled NLSE.

Once we establish the possibility of applying linear superposition principle to the two-coupled NLSE, we then show that the linear superposition, described above, is equivalent to a rotation in a two-dimensional function space where the two components of the seed solution are its coordinates. The rotation angle, $\theta$, turns out to be the only free parameter in this transformation. Taking the new solution as a seed and applying the transformation again, results in a new solution rotated by $2\theta$ with respect to the first seed solution. It will then be possible to write down a rotation matrix for a number $n$ of such consecutive rotations. This will result in a family composed of a series of solutions that are all generated from the same seed and rotation angle $\theta$. Changing the seed will result in a different family. However, changing the angle results in a different sequence of solutions within the same family. For instance, the rotation angle $\pi/2$, with a seed $(\psi_1,0)^T$, will result in the finite series of solutions:  $(0,\psi_2)^T$, $(-\psi_1,0)^T$, $(0,-\psi_2)^T$, and finally $(\psi_1,0)^T$, as a result of 4 iterations of the rotation operator. It is obvious that more rotations will not produce new solutions. This is a general property where angle of rotation satisfying $\theta=2\pi/n$ will result in a series of $n$ solutions. For other angles, an infinite series will be obtained. 

We then show that a linear superposition of two composite solutions, $(\Phi_1,\Phi_2)^T=d_1\,(\phi_1,\phi_2)^T + d_2\,(\phi_3,\phi_4)^T$, is indeed a solution. In view of the rotation interpretation,  the superposition can be viewed as another rotation operator. The solution $(\phi_1,\phi_2)^T$ on the right hand side of this expression is interpreted as a rotation of the seed  by an angle $\theta_1$, while $(\phi_3,\phi_4)^T$ is a rotation of the same seed by an angle $\theta_2$. The left hand side will be shown to correspond to a rotation of the same seed by an angle $\delta$ that may be given in terms of $\theta_1$ and $\theta_2$. As such, the superposition principle connects two solutions to a third solution, all of which belong to the same family generated by a specific seed solution. We have verified that the superposition principle is not applicable for seed solutions from different families.

As an application, we apply the superposition principle to almost all known solutions of the two-coupled NLSE. Most interestingly, the linear superposition of the localized solutions, such as the dark-bright soliton solution, leads to completely different type of solutions which are similar to breather solitons. Some other higher order solutions such as the Peregrine-bright soliton or Peregrine-dark soliton lead to a Peregrine-breather soliton. When the periodic solutions are considered, such as the Jacobi elliptic functions, the linear superposition results in a breather that is periodic in both space and time. This invokes the other three known breathers, namely the Akhmediev \cite{akhm}, Kusnetz-Ma \cite{km}, and Peregrine \cite{per} breathers, where the first is periodic in space and localized in time, the second is periodic in time and localized in space, and the third is localized in both space and time. The breather reported in this paper represents the fourth option, which is periodic in both space and time.

The superposition principle described here applies also to the $N$-coupled  NLSE. We show that explicitly  for $N=3$ with localized solutions, such as dark-dark-bright soliton and bright-bright-dark soliton as seed solutions.

The rest of the paper is organized as follows. In the next section, we first derive in Section \ref{gensub1} the transformation leading to composite solutions, show in Section \ref{gensub2} its analogy with a rotation operator, and then verify, in Section \ref{gensub3}, that the superposition of two composite solutions is also equivalent to a rotation operator. In Section \ref{secappl}, we apply the superposition principle to almost all known exact solutions of the two-coupled NLSE and generate new solutions. In Section \ref{secn3}, we consider the three-coupled NLSE. We end in Section \ref{concsec} by our main conclusions and outlook for future work.

\section{Composite solutions, rotation operator, and superposition principle}
\label{gensec}
In this section, we consider the two-coupled NLSE. We  first show that a linear combination of the two components of a given seed solution is also a solution which we denote as a composite solution. Then, we show that such kind of superposition is equivalent to a rotation operator in a two-dimensional function space with the components of the seed solution being its coordinates.  We show also that a repeated action of the rotation operator generates a finite or infinite series of solutions all belonging to one family characterized by a specific seed solution.  Finally, we consider the superposition of two composite solutions and show that it indeed satisfies the two-coupled NLSE and is also equivalent to a rotation in the function space.

\subsection{Composite solutions}
\label{gensub1}
Consider the two-coupled NLSE, alias vector NLSE or Manakov system \cite{man},
\begin{widetext}
\begin{eqnarray}
i\, \frac{\partial}{\partial t}\psi_1(x,t)+\frac{\partial^2}{\partial x^2}\psi_1(x,t)+\left(b_{11}\,|\psi_1(x,t)|^2+b_{12}\,|\psi_2(x,t)|^2\right)\,\psi_1(x,t)&=&0,\label{cnlse1}\\\label{cnlse2}
i\, \frac{\partial}{\partial t}\psi_2(x,t)+\frac{\partial^2}{\partial x^2}\psi_2(x,t)+\left(b_{21}\,|\psi_1(x,t)|^2+b_{22}\,|\psi_2(x,t)|^2\right)\,\psi_2(x,t)&=&0,
\end{eqnarray}
\end{widetext}
where, $\psi_1(x,t)$ and $\psi_2(x,t)$ are complex functions, $b_{11},\,b_{12},\,b_{21}$ and $b_{22}$ are arbitrary real constants representing the strengths of the nonlinear terms. In the context of Bose-Einstein condensates, this system models a two-component condensate, where $\psi_1(x,t)$ and $\psi_2(x,t)$ correspond to the componenets' wavefuncations \cite{chbook}. In another context, this system may model coupled modes in birefringent media or propagation of pulses in multi-mode fibers, where $\psi_1$ and $\psi_2$ represent the electric field intensity along each birefringence axis or fiber mode \cite{phbook}. We have initially considered a slightly more general system with different arbitrary dispersion coefficients, but it turned out that the superposition principle requires the two coefficients to be equal. In such a case, the coefficient of dispersion can be absorbed in a rescaling of $x$.

It is established that the system of Eqs. (\ref{cnlse1}) and (\ref{cnlse2}) is integrable and admits a Lax pair \cite{man}, and hence many of its exact solutions were found \cite{sols1,sols2,sols3,sols4,sols5,sols6,sols7,sols8,sols9,sols10}.  A list of all known solutions was recently complied in Ref.  \cite{ourbook}. We consider one of the exact solutions of  (\ref{cnlse1}) and (\ref{cnlse2}), which we denote by $\left(\psi_1,\psi_2\right)^T$ and refer to it as the seed solution.  Then, we construct  the following linear superposition of its components
\begin{eqnarray}
\phi_1(x,t)&=&c_1\,\psi_1(x,t)+c_2\,\psi_2(x,t)\label{ls1},\\
\phi_2(x,t)&=&c_3\,\psi_1(x,t)+c_4\,\psi_2(x,t),\label{ls2}
\end{eqnarray}
where $c_{1-4}$ are arbitrary real constants. We require  $\left(\phi_1,\phi_2\right)^T$ to be also a solution to (\ref{cnlse1})  and (\ref{cnlse2}) and denote it as a composite solution since each of its components is composed of a mix of the two components of the seed solution. Substituting (\ref{ls1}) and (\ref{ls2}) in (\ref{cnlse1}) and (\ref{cnlse2}) gives
\begin{eqnarray}
&&c_1 \left[i \psi _{1t}+\psi _{{1xx}}+\left({
   \left(b_{11} c_1^2+b_{12} c_3^2
   \right)|\psi _1|^2}+\frac{2 b_{11} c_1 c_2^2+b_{12} c_4 \left(c_2
   c_3+c_1 c_4\right)}{c_1}|\psi _2|^2
   \right)\psi _1+\frac{c_2  \left(b_{11} c_1 c_2+b_{12}
   c_3 c_4\right)}{c_1}\psi _2^2
  \psi_1^*\right]\nonumber\\
 &+& c_2 \left[i \psi _{2 t}+\psi _{2
   {xx}}+\left(\frac{
  2 b_{11} c_2 c_1^2+b_{12} c_3 \left(c_2
   c_3+c_1 c_4\right)}{c_2}|\psi _1|^2+{
    \left(b_{11} c_2^2+b_{12}
   c_4^2\right)}|\psi _2|^2\right)  \psi _2+\frac{c_1
   \left(b_{11} c_1 c_2+b_{12} c_3
   c_4\right)}{c_2}\psi_2^* \psi _1^2\right]=0\label{cnlse3},\nonumber\\\\\nonumber\\\nonumber\\
   &&c_3 \left[i \psi _{1t}+\psi _{{1xx}}+\left({
   \left(b_{22} c_3^2+b_{21} c_1^2
   \right)}|\psi _1|^2+\frac{
   2 b_{22} c_3 c_4^2+b_{21} c_2 \left(c_2
   c_3+c_1 c_4\right)}{c_3}|\psi _2|^2\right)\psi _1+\frac{c_4  \left(b_{21} c_1 c_2+b_{22}
   c_3 c_4\right)}{c_3}\psi _2^2
  \psi_1^*\right]\nonumber\\
  &+& c_4 \left[i \psi _{2 t}+\psi _{2
   {xx}}+\left(\frac{ 2 b_{22} c_4 c_3^2+b_{21} c_1 \left(c_2
   c_3+c_1 c_4\right)}{c_4}|\psi_1|^2+{
   \left(b_{22} c_4^2+b_{21}
   c_2^2\right)}|\psi _2|^2\right)\psi_2+\frac{c_3
   \left(b_{21} c_1 c_2+b_{22} c_3
   c_4\right)}{c_4}\psi_2^* \psi _1^2\right]=0\label{cnlse4},\nonumber\\
\end{eqnarray}
where, for convenience,  we hid the $x$- and $t$-dependence of $\psi_{1,2}$ and their complex conjugates and used subscripts to denote  partial derivatives with respect to $x$ and $t$. The expressions in the square brackets of Eqs. (\ref{cnlse3}) and (\ref{cnlse4}) correspond to the left hand side of Eqs. (\ref{cnlse1}) and (\ref{cnlse2}), but with the additional cross terms $\psi_2^2\psi_1^*$ and  $\psi_2^*\psi_1^2$. Requiring the cross terms to vanish and the constant coefficients of the other nonlinear terms in Eqs. (\ref{cnlse3}) and (\ref{cnlse4}) to match their counterparts in Eqs. (\ref{cnlse1}) and (\ref{cnlse2}) gives a set of equations for the unknown constants $c_{1-4}$ and imposes some restrictions on the strengths of the nonlinear terms, $b_{ij},\,\,i,j=1,2$. It turns out, however, that it is enough to match Eq. (\ref{cnlse3}) with Eqs. (\ref{cnlse1}) and (\ref{cnlse2}) since this guarantees the matching of Eq. (\ref{cnlse4}) to  Eqs. (\ref{cnlse1}) and (\ref{cnlse2}). Thus, the resulting equations for the coefficients are
\begin{equation}
b_{11}c_1c_2+b_{12}c_3c_4=0,
\end{equation}
\begin{equation}
b_{11}c_1^2+b_{12}c_3^2=b_{11},
\end{equation}
 \begin{equation}
\frac{2b_{11}c_1c_2^2+b_{12}c_4(c_2c_3+c_1c_4)}{c_1}=b_{12},
\end{equation}
 \begin{equation}
\frac{2b_{11}c_1^2c_2+b_{12}c_3(c_2c_3+c_1c_4)}{c_2}=b_{21},
\end{equation}
\begin{equation}
b_{11}c_2^2+b_{12}c_4^2=b_{22}.
\end{equation}
A nontrivial solution of this system of five equations gives three out of the coefficients $c_{1-4}$ in terms of one of them, which we choose to be $c_3$, and determines two of the nonlinear coefficients, $b_{ij},\,\,i,j=1,2$, in terms of the other two, which we choose to be $b_{11}$ and $b_{12}$
\begin{equation}
c_1=p\frac{\sqrt{b_{11}-b_{12}c_3^2}}{\sqrt{b_{11}}},\hspace{.5cm}
c_2=-pq\frac{b_{12}c_3}{b_{11}},\hspace{.5cm}
c_4=q c_1,
\end{equation} 
\begin{equation}
b_{21}=b_{11},\hspace{.5cm}b_{22}=b_{12},
\end{equation}
where $p=\pm1$ and $q=\pm1$ are independent from each other. The  specific choice of $p=q=1$ will be used for the rest of the paper. This choice does not affect the main features of the new solutions and it allows for satisfying a desired initial condition on the rotation operator of the next section.

Taking these solutions and restrictions into consideration, the two-coupled NLSE, (\ref{cnlse1}) and (\ref{cnlse2}), takes the form 
\begin{widetext}
\begin{eqnarray}
i\, \frac{\partial}{\partial t}\phi_1(x,t)+\frac{\partial^2}{\partial x^2}\phi_1(x,t)+\left(b_{11}\,|\phi_1(x,t)|^2+b_{12}\,|\phi_2(x,t)|^2\right)\,\phi_1(x,t)&=&0,\label{cnlse5}\\\label{cnlse6}
i\, \frac{\partial}{\partial t}\phi_2(x,t)+\frac{\partial^2}{\partial x^2}\phi_2(x,t)+\left(b_{11}\,|\phi_1(x,t)|^2+b_{12}\,|\phi_2(x,t)|^2\right)\,\phi_2(x,t)&=&0,
\end{eqnarray}
\end{widetext}
which supports the composite solution
\begin{eqnarray}
\phi_1(x,t)&=&{\sqrt{1-\frac{b_{12}}{b_{11}}c_3^2}}\,\psi_1(x,t)-\frac{b_{12}c_3}{b_{11}}\,\psi_2(x,t)\label{ls3},\\
\phi_2(x,t)&=&c_3\,\psi_1(x,t)+ {\sqrt{1-\frac{b_{12}}{b_{11}}c_3^2}}\,\psi_2(x,t),\label{ls4}
\end{eqnarray}
where, $\psi_{1,2}(x,t)$ are the two components of the seed solution satisfying Eqs. (\ref{cnlse5}) and (\ref{cnlse6}), upon substituting $\psi_1$ for $\phi_1$ and $\psi_2$ for $\phi_2$.

Thus, a composite solution can be constructed from an arbitrary mix of the two components of the seed solution, as determined by the last two equations.  However, we will argue below that it will be more convenient to re-express the composite solution in terms of a rotated seed solution in a two-dimensional function space.  

Being determined only by one arbitrary parameter, namely $c_3$, is a restriction imposed by the fact that the composite solution preserves the total norm of the two components. The arbitrariness is just in the relative percentages of each seed component. This suggests a geometrical interpretation to the composite solution as a rotation in a two-dimensional function space, which is the subject of the next section.

\subsection{Rotation operator}
\label{gensub2}
In this section, we show that the linear superposition that gives the composite solution (\ref{ls3}) and (\ref{ls4}) is equivalent to a rotation of the seed solution in a two-dimensional function space with components of the seed solution being its coordinates. 

We rewrite the only arbitrary parameter in the composite solution, namely $c_3$, as
\begin{equation}
c_3=-r\sin{\theta},
\end{equation}
\begin{equation}
r=\sqrt{\frac{b_{11}}{b_{12}}},
\end{equation}
where $\theta$ is an arbitrary real constant. With these substitutions, the composite solution (\ref{ls3}) and (\ref{ls4}) is written in vector form as
\begin{equation}
\left(\begin{array}{cc}\phi_1\\\phi_2\end{array}\right)=\left(\begin{array}{cc}\cos\theta&\frac{\sin\theta}{r}\\-r\sin\theta & \cos\theta\end{array}\right)\left(\begin{array}{cc}\psi_1\\\psi_2\end{array}\right)\label{rot}.
\end{equation}

\begin{figure}[!h]
	\centering
	\includegraphics[scale=1]{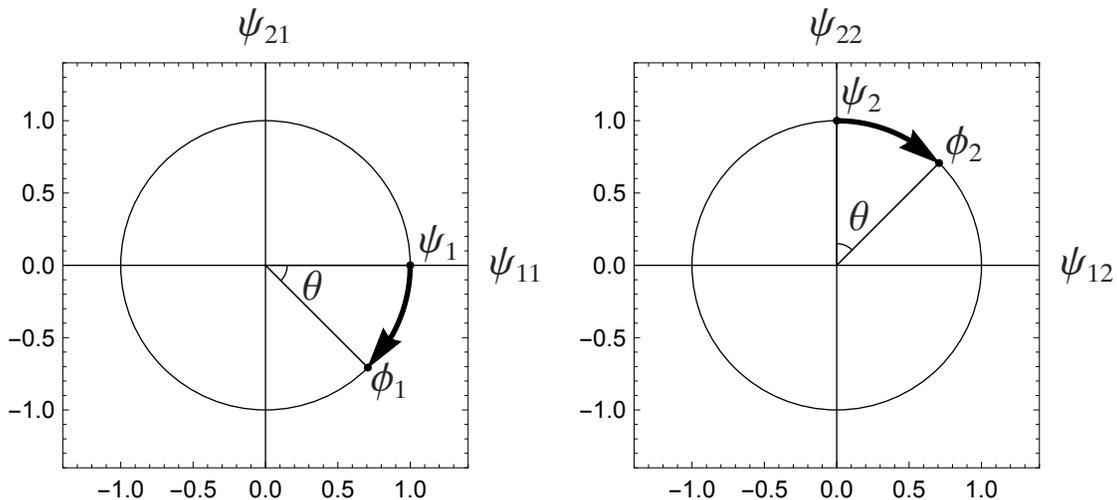}
	\caption{Schematic figure representing the rotation of the components of the seed solution $\psi=(\psi_1,\psi_2)^T$ by an angle $\theta=\pi/4$ to obtain the composite solution $\phi=(\phi_1,\phi_2)^T$, as given by Eq. (\ref{rot}). Left: rotation of $\psi_1$ with $\psi_{11}=1$ and $\psi_{21}=0$ to give $\phi_1=(1/\sqrt{2},-1/\sqrt{2})^T$, according to Eq. (\ref{rot4}). Right:  rotation of $\psi_2$ with $\psi_{12}=0$ and $\psi_{22}=1$ to give $\phi_2=(1/\sqrt{2},1/\sqrt{2})^T$, according to Eq. (\ref{rot5}). The composite solution  is obtained by combining the results of these two rotations, as given by Eqs. (\ref{phirot00}) and (\ref{phirot0}), namely $(\phi_1,\phi_2)^T=\left((\psi_1+\psi_2)/\sqrt{2},(-\psi_1+\psi_2)/\sqrt{2}\right)^T$.}
	\label{fig1}
\end{figure}

For $r=1$, the matrix on the right hand side of the last equation becomes the rotation matrix. This suggests an analogy with rotation of vectors in a two-dimensional plane. However, one should keep in mind that each of the components of the composite solution is a linear combination of the two components of the seed solution. Therefore, the appropriate representation of a composite solution is a two-dimensional second-rank tensor. To construct the tensor, we write the composite solution in the form
\begin{eqnarray}\label{phirot00}
\phi_1&=&(\phi_{11})\,\psi_1+(\phi_{21})\,\psi_2,\\
\phi_2&=&(\phi_{12})\,\psi_1+(\phi_{22})\,\psi_2,
\label{phirot0}
\end{eqnarray}
such that $\phi_{ij},\,\,j=1,2$ is the coefficient of $\psi_i$ in the $\phi_j$ component. 
The tensor is then written as
\begin{equation}
\phi=\left(\begin{array}{c}\phi_1\\\phi_2\end{array}\right)\Leftrightarrow\left(\begin{array}{cc}\phi_{11}&\phi_{12}\\ \phi_{21}&\phi_{22}\end{array}\right)\label{phirot},
\end{equation}
where the first column of this tensor corresponds to the coefficients of $\psi_1$ and $\psi_2$ in $\phi_1$ and the second column corresponds to the coefficients of $\psi_1$ and $\psi_2$  in $\phi_2$. The seed solution can also be formally represented by the tensor 
\begin{equation}
\psi=\left(\begin{array}{c}\psi_1\\\psi_2\end{array}\right)\Leftrightarrow\left(\begin{array}{cc}\psi_{11}&\psi_{12}\\ \psi_{21}&\psi_{22}\end{array}\right)\label{psirot}.
\end{equation}
Substituting the tensor forms of $\psi$ and $\phi$ from Eqs. (\ref{phirot}) and (\ref{psirot}) in Eq. (\ref{rot}), gives
\begin{widetext}
\begin{equation}
\left(\begin{array}{cc}\phi_{11}&\phi_{12}\\ \phi_{21}&\phi_{22}\end{array}\right)=\left(\begin{array}{cc}(\cos\theta)\psi_{11}+(\sin\theta)\psi_{21}&(\cos\theta)\psi_{12}+(\sin\theta)\psi_{22}\\(-\sin\theta)\psi_{11} +(\cos\theta)\psi_{21}& (-\sin\theta)\psi_{12} +(\cos\theta)\psi_{22}\end{array}\right)\label{rot2}.
\end{equation}
\end{widetext}
The first column of this equation can be written as
\begin{equation}
\left(\begin{array}{cc}\phi_{11}\\\phi_{21}\end{array}\right)=\left(\begin{array}{cc}\cos\theta&{\sin\theta}\\-\sin\theta & \cos\theta\end{array}\right)\left(\begin{array}{cc}\psi_{11}\\\psi_{21}\end{array}\right)\label{rot4},
\end{equation}
while the second column reads
\begin{equation}
\left(\begin{array}{cc}\phi_{12}\\\phi_{22}\end{array}\right)=\left(\begin{array}{cc}\cos\theta&{\sin\theta}\\-\sin\theta & \cos\theta\end{array}\right)\left(\begin{array}{cc}\psi_{12}\\\psi_{22}\end{array}\right)\label{rot5}.
\end{equation}
These two equations show that each component of the composite solution can be mapped to a rotation of a two-dimensional vector. For the first case, the components of the initial vector, which corresponds to $\psi_1$, are given by $(\psi_{11},\psi_{21})$ and the components of the rotated vector, which corresponds to $\phi_1$, are given by $(\phi_{11},\phi_{21})$. A similar analogy applies for the $\psi_2$ and $\phi_2$ components. The two rotations are depicted schematically in Fig. \ref{fig1}.

The composite solution $(\phi_1,\phi_2)^T$ obtained by the rotation in (\ref{rot}) can be used as a seed to the same rotation operator and another composite solution will be generated. Repeating this operation $n$ times generates the general solution
\begin{eqnarray}
\left(\begin{array}{cc}\phi_1\\\phi_2\end{array}\right)&=&\left(\begin{array}{cc}\cos \theta&\frac{\sin \theta}{r}\\-r\,\sin \theta & \cos \theta\end{array}\right)^n\left(\begin{array}{cc}\psi_1\\\psi_2\end{array}\right)\nonumber\\
&=&\left(\begin{array}{cc}\cos n\theta&\frac{\sin n\theta}{r}\\-r\,\sin n\theta & \cos n\theta\end{array}\right)\left(\begin{array}{cc}\psi_1\\\psi_2\end{array}\right)\label{rot6}.
\end{eqnarray}
For angles of rotation $\theta=2\pi/m$, where $m$ is an integer, a series of finite number of $m$ composite solutions will be obtained. The repeated action of the rotation operator for $n>m$ regenerates these solutions in a cyclic manner. For other values of $\theta$, an infinite series of composite solutions will be obtained. As and example, we consider the case of  repeated rotation by an angle $\theta=\pi/4$ which will generate the following series of 8 composite solutions, starting with the seed: $(\psi_1,\psi_2)^T$, $\left((\psi_1+\psi_2)/\sqrt{2},(-\psi_1+\psi_2)/\sqrt{2}\right)^T$, $(\psi_2,-\psi_1)^T$, $\left((-\psi_1+\psi_2)/\sqrt{2},-(\psi_1+\psi_2)/\sqrt{2}\right)^T$, $(-\psi_1,-\psi_2)^T$, $\left(-(\psi_1+\psi_2)/\sqrt{2},(\psi_1-\psi_2)/\sqrt{2}\right)^T$, $(-\psi_2,\psi_1)^T$, $\left((\psi_1-\psi_2)/\sqrt{2},(\psi_1+\psi_2)/\sqrt{2}\right)^T$. This sequence can be regenerated from the components of the tensor given by Eqs. (\ref{rot4}) and (\ref{rot5}) using the seed components  $(\psi_{11},\psi_{21})^T=(1,0)^T$ and $(\psi_{12},\psi_{22})^T=(0,1)^T$ . For instance, the application of the rotation in Eq. (\ref{rot4}) with $\theta=\pi/4$ gives $(\phi_{11},\phi_{21})^T=(1/\sqrt{2},1/\sqrt{2})^T$ and similarly Eq. (\ref{rot5}) gives $(\phi_{12},\phi_{22})^T=(-1/\sqrt{2},1/\sqrt{2})^T$. Combining these two results to obtain $\phi_1$ and $\phi_2$, as given by Eqs. (\ref{phirot0}) and (\ref{phirot00}), leads finally to $(\phi_1,\phi_2)^T=\left((\psi_1+\psi_2)/\sqrt{2},(-\psi_1+\psi_2)/\sqrt{2}\right)^T$, which corresponds to the first composite solution in the above-mentioned list.

Since for $r=1$ the rotation matrix preserves the norm, all composite solutions in this case will have their total norm squared equal to that of the seed solution
\begin{equation}
|\phi_1|^2+|\phi_2|^2=|\psi_1|^2+|\psi_2|^2.
\end{equation}

The representation of the composite solutions as given by Eq. (\ref{rot}) will be used in the following sections to generate new solutions out of the superposition of known ones.

\subsection{Superposition principle}
\label{gensub3}
In this section, we prove that the superposition of two composite solutions is also a composite solution to the two-coupled NLSE. To that end, we will employ the rotation representation described in the previous section. Consider a composite solution obtained by the rotation of a given seed by an angle $\theta_1$
\begin{equation}
\left(\begin{array}{c}\Psi_1\\\Psi_2\end{array}\right)=\left(\begin{array}{cc}\cos\theta_1&{\sin\theta_1}\\-\sin\theta_1 & \cos\theta_1\end{array}\right)\left(\begin{array}{cc}\psi_1\\\psi_2\end{array}\right)\label{Psieq}
\end{equation}
and another composite solution obtained by rotating the same seed by an angle $\theta_2=\theta_1+\phi$
\begin{equation}
\left(\begin{array}{c}\Phi_1\\\Phi_2\end{array}\right)=\left(\begin{array}{cc}\cos\theta_2&\sin\theta_2\\-\sin\theta_2 & \cos\theta_2\end{array}\right)\left(\begin{array}{cc}\psi_1\\\psi_2\end{array}\right)\label{Phieq}.
\end{equation}
The superposition 
\begin{equation}
\left(\begin{array}{c}\Omega_1\\\Omega_2\end{array}\right)=d_1\,\left(\begin{array}{c}\Psi_1\\\Psi_2\end{array}\right)+d_2\,\left(\begin{array}{c}\Phi_1\\\Phi_2\end{array}\right)\label{super}
\end{equation}
is required to satisfy the two-coupled NLSE (\ref{cnlse5}) and (\ref{cnlse6}), with $\phi_1=\Omega_1$ and $\phi_2=\Omega_2$, which imposes the following condition on the arbitrary coefficients $d_1$ and $d_2$
\begin{equation}
d_1^2+d_2^2+2d_1d_2\,\cos\phi=1\label{cond}.
\end{equation}
Therefore, the superposition principle applies to the composite solutions but with the two arbitrary constants $d_1$ and $d_2$ being reduced to one since they are related to each other as in the last equation.

It can be shown that the superposition of the two composite solutions, (\ref{super}), is equivalent to a rotation of the same seed solution by an angle $\delta$ as follows. Substituting for the composite solutions $(\Psi_1,\Psi_2)^T$ and $(\Phi_1,\Phi_2)^T$ from Eqs. (\ref{Psieq}) and (\ref{Phieq}) in Eq.(\ref{super}), gives
\begin{equation}
\left(\begin{array}{c}\Omega_1\\\Omega_2\end{array}\right)=\left(\begin{array}{cc}d_1\cos\theta_1+d_2\cos(\theta_1+\phi)&d_1\sin\theta_1+d_2\sin(\theta_1+\phi)\\\-d_1\sin\theta_1-d_2\sin(\theta_1+\phi)&d_1\cos\theta_1+d_2\cos(\theta_1+\phi)\end{array}\right)\left(\begin{array}{c}\psi_1\\\psi_2\end{array}\right),\label{super2}
\end{equation}
which can be put in the form of a rotation matrix
\begin{equation}
\left(\begin{array}{c}\Omega_1\\\Omega_2\end{array}\right)=\left(\begin{array}{cc}\cos\delta&\sin\delta\\-\sin\delta&\cos\delta\end{array}\right)\left(\begin{array}{c}\psi_1\\\psi_2\end{array}\right)\label{super3}
\end{equation}
with $d_1=-\sin(\delta-\theta_1-\phi)/\sin\phi$ and $d_2=\sin(\delta-\theta_1)/\sin\phi$ that also satisfy the condition (\ref{cond}). The $\phi=0$ case should be treated separately by setting $\phi=0$ in Eqs. (\ref{cond}) and (\ref{super2}) and then finding $d_1$ and $d_2$ by comparing the resulting matrix with Eq. (\ref{super3}). 

Equation (\ref{super2})  proves that the superposition of two composite solutions is also a composite solution with all three solutions  belonging to the same seed. In the next section, we will apply the superposition principle to almost all known solutions of the two-coupled NLSE to obtain composite solutions with some features that do not exist in the seed solutions.

\section{Composite solutions to the two-component Manakov system }
\label{secappl}
In this section we will use almost all the known solutions to the two-coupled NLSE as a seed to generate composite solutions using the rotation operator (\ref{rot}). A list of all known solutions to the two-coupled NLSE can be found in \cite{ourbook}. In the following, we will consider these solutions one by one, derive their composite solution, and then point out the main features incorporated by the superposition. Both the seed solution, $(\psi_1,\psi_2)^T$, and the composite solution, $(\phi_1,\phi_2)^T$, satisfy the two-coupled NLSE, Eqs. (\ref{cnlse1}) and (\ref{cnlse2}) and Eq. (\ref{cnlse5}) and (\ref{cnlse6}), respectively.\\

\begin{figure}[!h]
	\centering
	\includegraphics[scale=1]{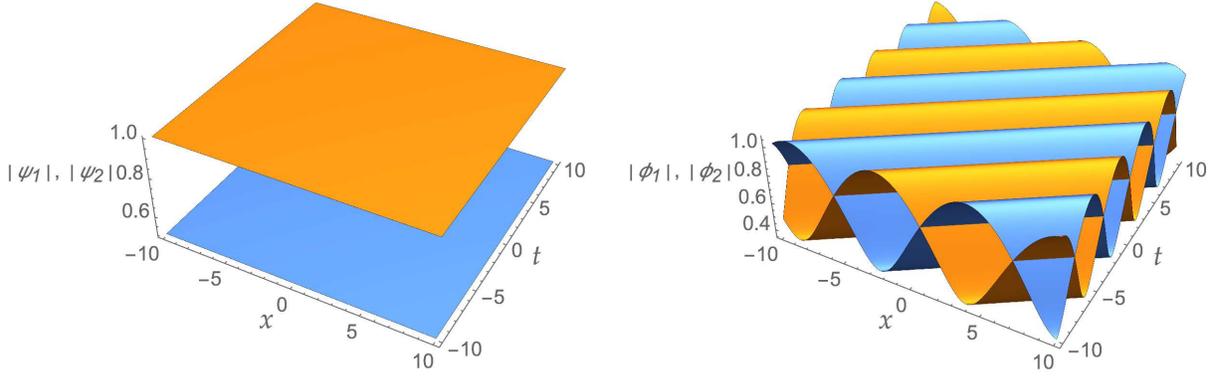}
	\caption{The norm of the CW seed solution (left) and composite solution (right), given by Eqs. (\ref{seed1}) and (\ref{comp1}). Orange corresponds to $\psi_1$ and $\phi_1$ and blue corresponds to $\psi_2$ and $\phi_2$. Parameters used: $A_0=A_2=b_{11}=b_{12}=1$, $B_0=B_2=1/2$, $\varphi_1=\varphi_2=0$, and $\theta=\pi/4$ for the composite solution.}
	\label{fig2}
\end{figure}

{\it Solution 1. Constant Wave  (CW):} Applying the rotation operator (\ref{rot}) on the seed solution 
\begin{equation}\label{seed1}
\left(\begin{array}{c}\psi_1\\\psi_2\end{array}\right)=\left(\begin{array}{cc}A_0\,e^{i\,(A_1t+A_2x+\varphi_1)}\\B_0\,e^{i(B_1t+B_2x+\varphi_2)}\end{array}\right)
\end{equation} with $A_1=-A_2^2+A_0^2\,b_{11}+B_0^2b_{12}$, $B_1=-B_2^2+A_0^2\,b_{11}+B_0^2b_{12}$, and $A_0,\,A_2,\,B_0,\,B_2,\,\varphi_1,\,\varphi_2$ are arbitrary real constants, will lead to the composite solution
\begin{widetext}
\begin{equation}\label{comp1}
\left(\begin{array}{c}\phi_1\\\phi_2\end{array}\right)=\left(\begin{array}{cc}A_0\,e^{i\,(A_1t+A_2x+\varphi_1)}\,\cos\theta+B_0\sqrt{\frac{b_{12}}{b_{11}}}\,e^{i(B_1t+B_2x+\varphi_2)}\,\sin\theta\\-A_0\sqrt{\frac{b_{11}}{b_{12}}}\,e^{i\,(A_1t+A_2x+\varphi_1)}\,\sin\theta+B_0\,e^{i(B_1t+B_2x+\varphi_2)}\,\cos\theta\end{array}\right).
\end{equation} 
\end{widetext}
The effect of the superposition is to introduce time-dependence to the norm of the composite solution that did not exist in the seed solution. While the norm squared of the seed solution components $|\psi_1|^2=A_0^2$ and $|\psi_2|^2=B_0^2$ are time-independent, the norm squared of the composite solution
\begin{widetext}
\begin{equation}
2b_{11}|\phi_1|^2=2A_0^2b_{11}\cos^2\theta+2B_0^2b_{12}\sin^2\theta+2A_0B_0\sqrt{b_{11}b_{12}}\,\cos{\left[(A_2-B_2)\left((A_2+B_2)t-x\right)+\varphi_2-\varphi_1\right]}\,\sin{(2\theta)},
\end{equation} 
\begin{equation}
2b_{12}|\phi_2|^2=2A_0^2b_{11}\sin^2\theta+2B_0^2b_{12}\cos^2\theta-2A_0B_0\sqrt{b_{11}b_{12}}\,\cos{\left[(A_2-B_2)\left((A_2+B_2)t-x\right)+\varphi_2-\varphi_1\right]}\,\sin{(2\theta)},
\end{equation}
\end{widetext} 
may have time-dependence for $\theta\ne\pi/2$. Clearly, these expressions give the norms of the seed solution for $\theta=0$, but for example, when $\theta=\pi/4$, the last term in both equations will introduce a sinusoidal time variation of the background which, in this case, becomes traveling waves with speed $A_2^2-B_2^2$. In Fig. \ref{fig2}, we plot the norms of the seed solution and the composite solution for $\theta=\pi/4$. The figure shows that the norms of the composite solution are out-of-phase traveling waves.\\

\begin{figure}[!h]
	\centering
	\includegraphics[scale=1]{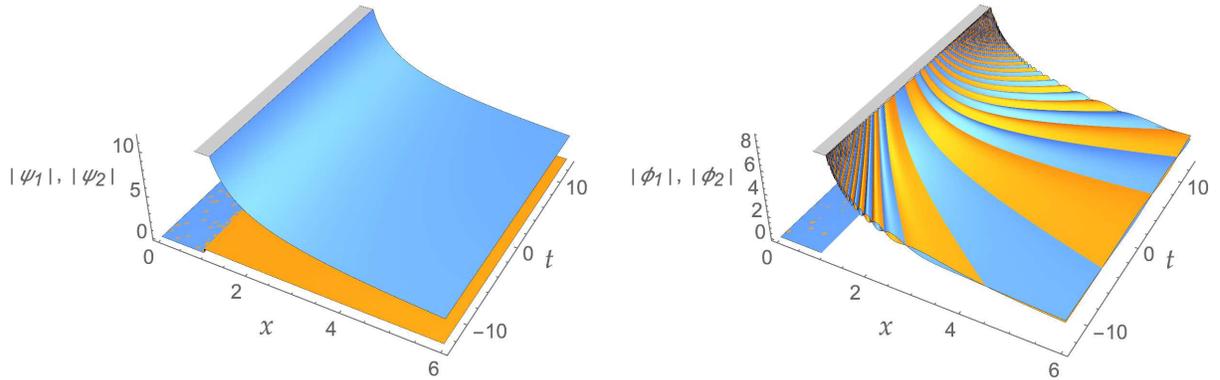}
	\caption{The norm of seed {Solution 2}  (left) and its composite solution (right), given by Eqs. (\ref{seed21})-(\ref{comp22}), respectively. Orange corresponds to $\psi_1$ and $\phi_1$ and blue corresponds to $\psi_2$ and $\phi_2$. Parameters used: $A_0=1, B_0=10, A_1=1, B_1=-1, A_2=-2, B_2=2, b_1=b_2=1$, $\varphi_1=\varphi_2=0$, and $\theta=\pi/4$ for the composite solution.}
	\label{fig3}
\end{figure}

{\it Solution 2. Rational solution: decaying wave}.  
The seed solution is given by
\begin{eqnarray}\label{seed21}
\psi_1&=&\frac{A_0}{\sqrt{2(A_1+t)}}\\\nonumber&&\times e^{i\,\left[\frac{(A_2+x)^2}{4(A_1+t)}+\frac{A_0^2b_{11}}{2}\log(2(A_1+t))+\frac{B_0^2b_{12}}{2}\log(2(B_1+t))+\varphi_1\right]},\\
\psi_2&=&\frac{B_0}{\sqrt{2(B_1+t)}}\nonumber\\&&\times e^{i\,\left[\frac{(B_2+x)^2}{4(B_1+t)}+\frac{A_0^2b_{11}}{2}\log(2(A_1+t))+\frac{B_0^2b_{12}}{2}\log(2(B_1+t))+\varphi_2\right]}\label{seed22},
\end{eqnarray}
where $A_0,\,B_0\,A_1\,B_1,\,A_2,\,B_2,\,\varphi_1,\,\varphi_2$ are arbitrary real constants. Constructing the composite solution by applying the rotation operator (\ref{rot}), and then calculating the square norm, we get
\begin{widetext}
\begin{eqnarray}\label{comp21}
|\phi_1|^2&=&\frac{1}{2b_{11}t}\left(A_0^2b_{11}\cos^2\theta+B_0^2b_{12}\sin^2\theta+A_0B_0\sqrt{b_{11}b_{12}}\,\cos{\left(\frac{A_2(A_2+2x)-B_2(B_2+2x)+4t(\varphi_1-\varphi_2)}{4t}\right)}\,\sin(2\theta)\right),\nonumber\\\\\label{comp22}
|\phi_2|^2&=&\frac{1}{2b_{12}t}\left(A_0^2b_{11}\sin^2\theta+B_0^2b_{12}\cos^2\theta-A_0B_0\sqrt{b_{11}b_{12}}\,\cos{\left(\frac{A_2(A_2+2x)-B_2(B_2+2x)+4t(\varphi_1-\varphi_2)}{4t}\right)}\,\sin(2\theta)\right),\nonumber\\
\end{eqnarray}
\end{widetext}
where we have set  $A_1=B_1=0$, which correspond to the time reference and does not affect the main feature of the composite solution. Here, again, the superposition introduces a traveling wave on top of the seed solution decaying background, as shown in Fig. \ref{fig3}.\\

\begin{figure}[!h]
	\centering
	\includegraphics[scale=1]{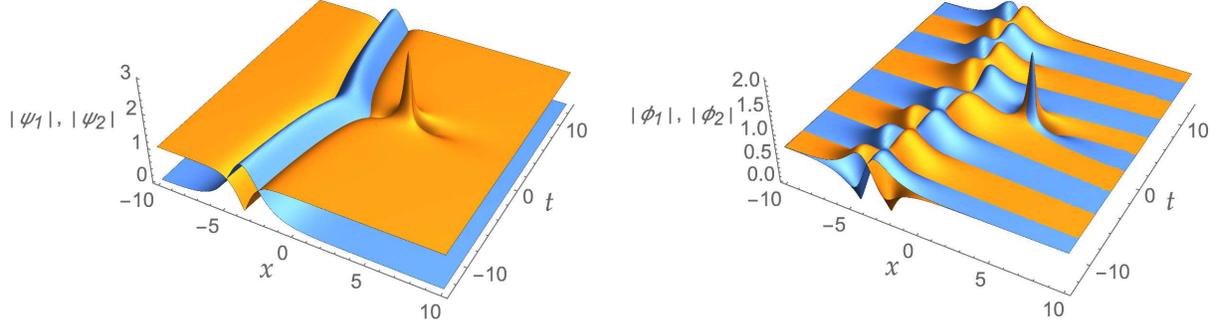}
	\caption{The norm of seed {Solution 3}  (left) and its composite solution (right). Orange corresponds to $\psi_1$ and $\phi_1$ and blue corresponds to $\psi_2$ and $\phi_2$. Parameters used: $A_0=-1, A_1=0, \beta_1=-3, \beta_2=\beta_3=0, \beta_4=1$, and $, \theta=\pi/4$ for the composite solution.}
\label{fig4}
\end{figure}

{\it Solution 3. Bright-dark soliton scattered by a Peregrine soliton}.
This seed solution corresponds to a dark (bright) soliton and a Perigrine soliton in one component and a bright (dark) soliton in the other component. The solution is given by
\begin{eqnarray}
\psi_1&=&A_0\,e^{i\left(A_1x+\left(2A_0^2-A_1^2\right)t\right)}\nonumber\\&&\times \left[1-\frac{4e^{-\eta}\left(\alpha_1^2+\alpha_2^2-\alpha_1+i\,\alpha_2\right)}{e^{2\eta}(\beta_3^2+\beta_4^2)+e^{-\eta}\left(1+2\alpha_1^2+2\alpha_2^2-2\alpha_1\right)}\right]\label{seed3},
\end{eqnarray} 
\begin{eqnarray}
\psi_2&=&-4A_0\,e^{i\left(A_1(x-A_1t)+3A_0^2t\right)+3\eta/2}\nonumber\\&&\times \frac{\left(\beta_3+i\beta_4\right)\left(\alpha_1+i\alpha_2-1\right)}{1+e^{3\eta}(\beta_3^2+\beta_4^2)+2 e^{-\eta}\left(\alpha_1-1\right)\alpha_1+2\alpha_2^2}\label{seed3},
\end{eqnarray}
where $\alpha_1(x,t)=\beta_1A_0+A_0(x-2A_1t)$, $\alpha_2(x,t)=\beta_2A_0-2A_0^2t$, $\eta(x,t)=2A_0(x-2A_1t)/3$, $b_{11}=b_{12}=b_{21}=b_{22}=2$, and $A_0, A_1, \beta_{1-4}$ are arbitrary real constants. Obtaining the composite solution obtained by applying the rotation operator (\ref{rot}) is straight forward but generates lengthy expressions that we do not present here. Figure \ref{fig4} shows that the $\psi_1$ component of the seed solution is  a dark-Peregrine soliton while the $\psi_2$ component is a bright soliton. The figure indicates an attractive force of interaction between the dark and bright solitons on one hand and the Peregrine soliton on the other hand. The corresponding composite solution with $\theta=\pi/4$ is also shown in this figure where the bright and dark solitons turn to out-of-phase breathers. This is a fundamentally different feature which is not present in the seed solution.\\

\begin{figure}[!h]
	\centering
	\includegraphics[scale=1]{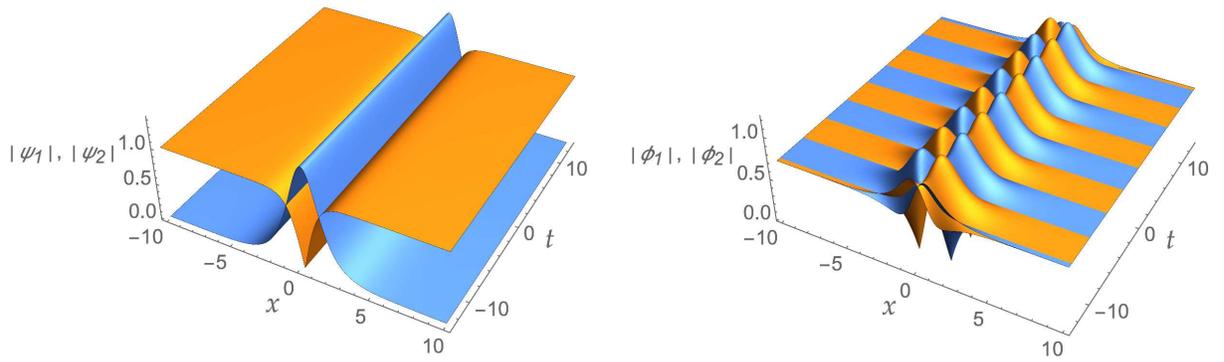}
	\caption{The norm of seed {Solution 4}  (left) and its composite solution (right), given by Eqs. (\ref{seed41})-(\ref{comp42}). Orange corresponds to $\psi_1$ and $\phi_1$ and blue corresponds to $\psi_2$ and $\phi_2$. Parameters used: $A_0=1, A_1=0$, and $, \theta=\pi/4$ for the composite solution.}
\label{fig5}
\end{figure}

{\it Solution 4. Bright-dark soliton:} This solution is a limiting case of Solution 3, with $\beta_1=\beta_4$ and $\beta_4\rightarrow\infty$. In this limit, the Peregrine soliton disappears leaving the bright soliton in one component and a dark soliton in the other component. The seed solution simplifies in this case to
\begin{equation}
\psi_1=A_0\, e^{2iA_0^2t}{\rm tanh}(A_0 x)\label{seed41},
\end{equation}
\begin{equation}
\psi_2=-i\sqrt{2}\,A_0\, e^{3iA_0^2t}{\rm sech}(A_0 x)\label{seed42},
\end{equation}
where $A_0$ is an arbitrary real constant. The rotation operator (\ref{rot}) gives its composite solution as
\begin{equation}
\phi_1=A_0\, e^{2iA_0^2t}\left[{\rm tanh}(A_0 x)\,\cos\theta-i\sqrt{2}\, e^{iA_0^2t}{\rm sech}(A_0 x)\,\sin\theta\right]\label{comp41},
\end{equation}
\begin{equation}
\phi_2=-A_0\, e^{2iA_0^2t}\left[{\rm tanh}(A_0 x)\,\sin\theta+i\sqrt{2}\, e^{iA_0^2t}{\rm sech}(A_0 x)\,\cos\theta\right]\label{comp42}.
\end{equation}
While the seed solutions (\ref{seed41}) and (\ref{seed42}) are stationary solutions, the composite solutions (\ref{comp41}) and (\ref{comp42}) are not so due to the phase factor $e^{iA_0^2t}$ which will lead to time-dependent cross terms in the norm. The calculation of square norm of the two components of the composite solution shows this additional time-dependence most clearly
\begin{equation}
|\phi_1|^2=2A_0^2{\rm sech}^2(A_0x)\,\sin^2\theta+A_0^2{\rm tanh}^2(A_0x)\,\cos^2\theta+\sqrt{2}\,A_0^2{\rm sech}(A_0x)\,{\rm tanh}(A_0x)\,\sin(A_0^2t)\,\sin(2\theta)\label{norm41},
\end{equation}
\begin{equation}
|\phi_2|^2=2A_0^2{\rm sech}^2(A_0x)\,\cos^2\theta+A_0^2{\rm tanh}^2(A_0x)\,\sin^2\theta-\sqrt{2}\,A_0^2{\rm sech}(A_0x)\,{\rm tanh}(A_0x)\,\sin(A_0^2t)\,\sin(2\theta)\label{norm42}.
\end{equation}
For $\theta=0$, the square norm of the components of the seed solution will be retrieved. For $\theta\ne\pi/2$, the $\sin(A_0^2t)$ time-dependent term will give rise to an additional oscillatory behavior in the norms. This is shown in Fig. \ref{fig5} where a seed solution of stationary dark-bright soliton turns to composite solution of a breather dark-bright soliton oscillating with frequency $A_0^2$.\\

{\it Solution 5. Two Peregrine solitons:}  This higher order Peregrine soliton solution corresponds to a Peregrine soliton in each of the two components of the seed solution, which takes the form
\begin{equation}
\psi_1=A_0\,e^{i\left(16A_0^2t+A_1(x-A_1t)\right)}\left[-1-i\sqrt{3}+\frac{-3+5i\sqrt{3}-36(\sqrt{3}-i)A_0^2t-6(\sqrt{3}-i)A_0\,\eta}{5+144A_0^4t^2+4A_0\,\eta\,(2\sqrt{3}+3A_0\eta)}\right]\label{seed51},
\end{equation} 
\begin{equation}
\psi_2=A_0\,e^{i\left(16A_0^2t+A_2(x-A_2t)\right)}\left[-1+i\sqrt{3}+\frac{-3-5i\sqrt{3}+36(\sqrt{3}+i)A_0^2t-6(\sqrt{3}+i)A_0\,\eta}{5+144A_0^4t^2+4A_0\,\eta\,(2\sqrt{3}+3A_0\eta)}\right]\label{seed52},
\end{equation} 
where $\eta=x+6A_3t$, $A_1=A_2-2A_0$, $A_0=A_2+3A_3$, and $A_0, A_1, A_2, A_3$ are arbitrary real constants. The composite solution is lengthy and therefor will not be presented here.\\
\begin{figure}[!h]
	\centering
	\includegraphics[scale=1]{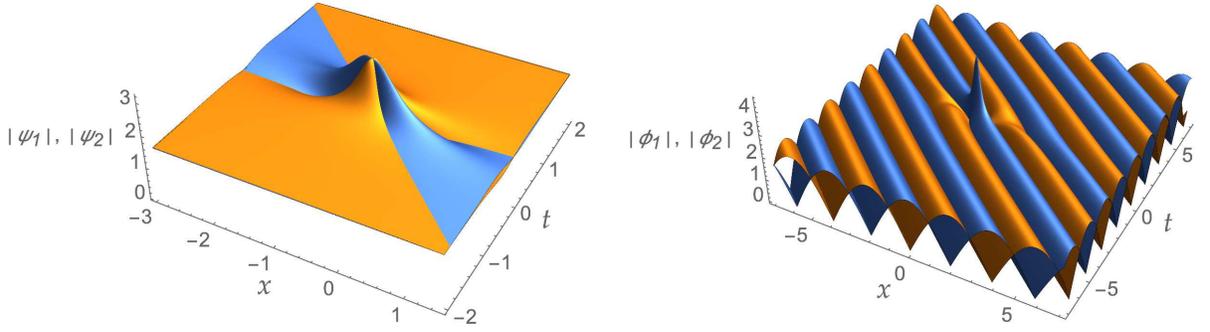}
	\caption{The norm of seed {Solution 5}  (left) and its composite solution (right). Orange corresponds to $\psi_1$ and $\phi_1$ and blue corresponds to $\psi_2$ and $\phi_2$. Parameters used: $A_2=A_3=0.2$, $b_{11}=b_{12}=b_{21}=b_{22}=2$, and $, \theta=\pi/4$ for the composite solution.}
\label{fig6}
\end{figure}

\begin{figure}[!h]
	\centering
	\includegraphics[scale=1]{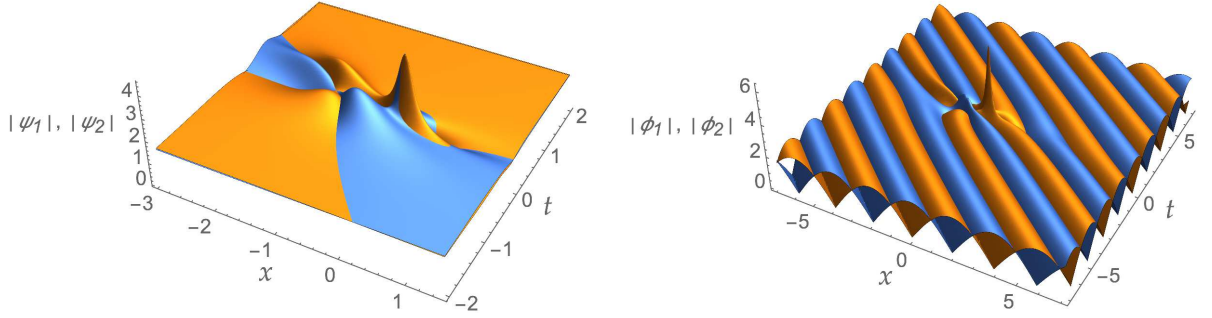}
	\caption{The norm of seed {Solution 6}  (left) and its composite solution (right). Orange corresponds to $\psi_1$ and $\phi_1$ and blue corresponds to $\psi_2$ and $\phi_2$. Parameters used: $A_2=A_3=0.2$, $b_{11}=b_{12}=b_{21}=b_{22}=2$, and $, \theta=\pi/4$ for the composite solution.}
\label{fig7}
\end{figure}

{\it Solution 6. Higher order Peregrine soliton:} This solution corresponds to a higher order Peregrine soliton. The lengthy expressions of $\psi_1$ and $\psi_2$ are relegated to Appendix A. In Fig. \ref{fig7}, we show the seed solution together with its composite solution for $\theta=\pi/4$, where here again an oscillatory background is being brought by the composite solution.\\

\begin{figure}[!h]
	\centering
	\includegraphics[scale=1]{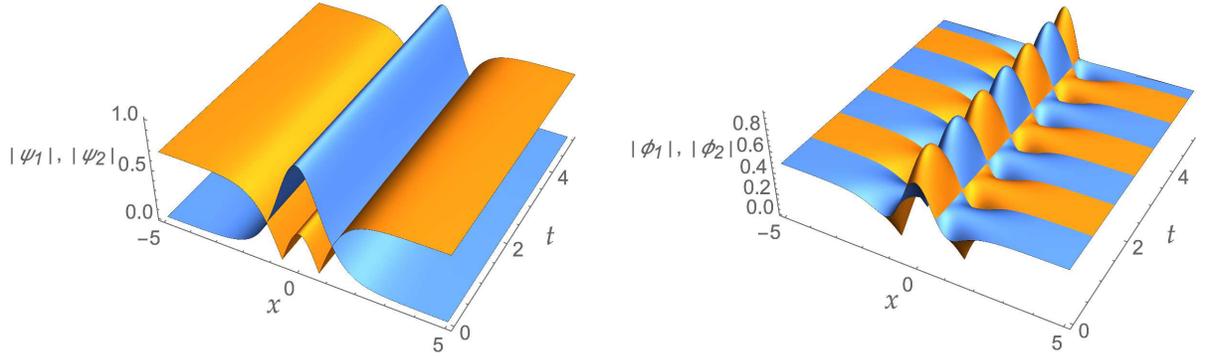}
	\caption{The norm of seed {Solution 7}  (left) and its composite solution (right), given by Eqs. (\ref{seed71})-(\ref{comp72}). Orange corresponds to $\psi_1$ and $\phi_1$ and blue corresponds to $\psi_2$ and $\phi_2$. Parameters used: $A_1=A_0=B_0=1$, and $\theta=\pi/4$ for the composite solution.}
\label{fig8}
\end{figure}

{\it Solution 7. ${\rm sech}^2$-solution:} This solution corresponds to two bright solitons with different backgrounds. It reads
\begin{equation}
\psi_1=e^{-i\left(\omega_1t+\varphi_1\right)}\,\left[A_0\,{\rm sech}^2(A_1x)+A_3\right]\label{seed71},
\end{equation}
\begin{equation}
\psi_2=B_0\,e^{-i\left(\omega_2t+\varphi_2\right)}\,{\rm sech}^2(A_1x)\label{seed72},
\end{equation}
where $A_3=-2A_0/3$, $\omega_1=2A_1^2$, $\omega_2=-2A_1^2$, $b_1=-9A_1^2/(2A_0^2)$, $b_2=9A_1^2/(2B_0^2)$, $b_{12}=b_{11}$, and $b_{21}=b_{22}$. The composite solution takes the form
\begin{eqnarray}
\phi_1&=&e^{-i\left(\omega_1t+\varphi_1\right)}\left[A_0\,{\rm sech}^2(A_1x)+A_3\right]\,\cos\theta\nonumber\\&&+iB_0\,e^{-i\left(\omega_2t+\varphi_2\right)}\,{\rm sech}^2(A_1x)\,\sin\theta
\label{comp71},
\end{eqnarray}
\begin{eqnarray}
\phi_2&=&B_0\,e^{-i\left(\omega_2t+\varphi_2\right)}\,{\rm sech}^2(A_1x)\,\cos\theta\nonumber\\&&-e^{-i\left(\omega_1t+\varphi_1\right)}\,\left[A_3+A_0\,{\rm sech}^2(A_1x)\right]\,\sin\theta
\label{comp72}.
\end{eqnarray}
Similar to Solution 4, one can calculate the norm of the components of the composite solution to show that an additional time-dependent oscillator term with prefactor $\sin(2\theta)$ will appear as a result of the superposition. This term gives rise to the oscillations in the background, as shown in Fig. \ref{fig8}. Here again, the oscillation in the $\phi_1$ component is out-of-phase with respect to that in the $\phi_2$ component.\\

\begin{figure}[!h]
	\centering
	\includegraphics[scale=1]{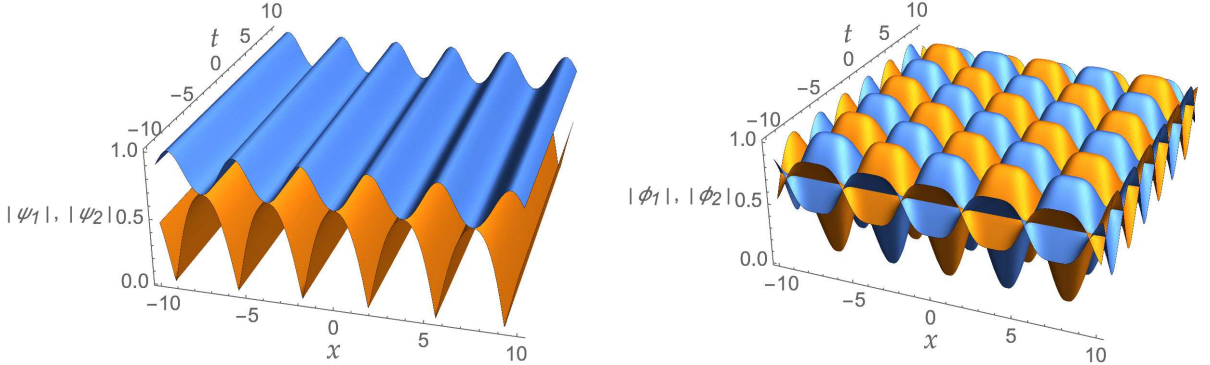}
	\caption{The norm of seed {Solution 8}  (left) and its composite solution (right), given by Eqs. (\ref{seed81})-(\ref{comp82}). Orange corresponds to $\psi_1$ and $\phi_1$ and blue corresponds to $\psi_2$ and $\phi_2$. Parameters used: $A_1=A_0=B_0=1, m=1/2$, and $\theta=\pi/4$ for the composite solution.}
\label{fig9}
\end{figure}

{\it Solution 8. Solitary waves:} This solution is a solitary wave given in terms of the Jacobi elliptic functions
\begin{equation}
\psi_1=A_0\,e^{-i(\omega_1t+\varphi_1)}\sqrt{m}\,{\rm cd}(A_1x|m)\label{seed81},
\end{equation}
\begin{equation}
\psi_2=B_0\,e^{-i(\omega_2t+\varphi_2)}\sqrt{1-m}\,{\rm nd}(A_1x|m)\label{seed82},
\end{equation}
where ${\rm cd}(x|m)$ and ${\rm nd}(x|m)$ are the Jacobi elliptic functions with modulus $k^2=m$, $\omega_1=-(1-m)A_1^2-b_{11}A_0^2$, $\omega_2=-(2-m)A_1^2-b_{11}A_0^2$, 
$b_{11}=(-2A_1^2+b_{12}B_0^2)/A_0^2$. The composite solution then reads
\begin{eqnarray}
\phi_1&=&A_0\sqrt{m}\,e^{-i\left((A_1^2(1+m)-B_0^2b_{12})t+\varphi_1\right)}{\rm cd}(A_1x|m)\cos\theta\nonumber\\&&+B_0\sqrt{1-m}\sqrt{\frac{A_0^2b_{12}}{B_0^2b_{12}-2A_1^2}}\,e^{-i\left((A_1^2m-B_0^2b_{12})t+\varphi_2\right)}{\rm nd}(A_1x|m)\,\sin\theta\label{comp81},
\end{eqnarray}
\begin{eqnarray}
\phi_2&=&-A_0\sqrt{m}\,\sqrt{\frac{B_0^2b_{12}-2A_1^2}{A_0^2b_{12}}}\,e^{-i\left((A_1^2(1+m)-B_0^2b_{12})t+\varphi_1\right)}{\rm cd}(A_1x|m)\sin\theta\nonumber\\&&+B_0\sqrt{1-m}\,e^{-i\left((A_1^2m-B_0^2b_{12})t+\varphi_2\right)}{\rm nd}(A_1x|m)\,\cos\theta\label{comp82}.
\end{eqnarray}
The seed solution is a stationary solitary wave such that its norm is independent of time. The composite solution may on the other hand have a norm that not only oscillates in space, but also oscillates in time. This is shown in Fig. \ref{fig9}. The periodicity of the composite solution in both space and time suggests an additional breather to the other three known breathers, namely the Akhmediev \cite{akhm} breather which is periodic in space and localized in time, Kusnetz-Ma \cite{km} breather which is periodic in time and localized in space, and the Peregrine \cite{per} breather which is localized in both space and time. 

With this, we have considered most of the known solutions to the two-coupled NLSE. There are at least two solutions, that we are aware of, which were not considered here. The first is a dark-bright solution similar to our Solution 4 and corresponds to Solution 6 in \cite{ourbook}. However, this solution requires the signs of the dispersion terms to be opposite. We have pointed out in section \ref{gensec} that for the superposition principle to be applicable, the two dispersion terms must have the same coefficient. Therefore, the superposition principle does not apply to this solution. The other solution that we did not consider is the Weierstrass elliptic function \cite{str}, which seemed to be more complicated than the solutions considered here and will have even much more complicated composite solution.

In principle, the idea of composite solutions can be generalized to $N$-coupled NLSE. In the next section, we generalize the idea of composite solutions to three-coupled NLSE and consider two specific examples.

\section{Composite solutions to the N-coupled nonlinear Schr\"odinger equation}
\label{secn3}
In this section, we consider the $N$-coupled NLSE \cite{sols4,sols8,sols9,sols11}, with emphasis on the three-coupled NLSE. Similar to the previous case of two-coupled NLSE, we show that a linear combination of the three components of a given  seed solution  is also a composite solution. In addition, we present a symmetry reduction that transforms the three-coupled NLSE to the two-coupled NLSE. Such symmetry reduction will be an efficient tool to obtain one-to-one correspondence  between three-coupled NLSE solutions and two-coupled NLSE solutions. 

Obtaining composite solutions to the three-coupled NLSE can thus be performed through one of two routs. The first rout is to apply the symmetry reduction to seed solutions of the two-coupled NLSE which leads to seed solutions of the three-coupled NLSE. Then we apply the superposition principle to the seed solution in order to obtain the composite solution of the three-coupled NLSE. The second rout is to apply the symmetry reduction to the composite solutions of the two-coupled NLSE. Then use the results as seed solutions of the three-coupled NLSE, and then apply the superposition principle to get their corresponding composite solutions.  We apply this symmetry reduction to two of the solutions presented above.

\subsection{Composite solutions}
The $N$-coupled NLSE is written as
\begin{equation}\label{ncnlse2}
i\, \frac{\partial}{\partial t}\psi_j(x,t)+\frac{\partial^2}{\partial x^2}\psi_{j}(x,t)+\left(\sum\limits_{k=1}^{N}\,{b_j}_k\,|\psi_k(x,t)|^2\right)\,\psi_j(x,t)=0,\hspace{1cm}j=1,\,2,\,\dots,\,N,
\end{equation}
where, $\psi_j(x,t)$ are complex functions and ${b_j}_k$ are arbitrary real constants. The exact solution to the system above, $(\psi_j)^T$, can be composed through the following linear superposition
\begin{equation}
\phi_j(x,t)=\sum\limits_{k=1}^{N}{c_j}_k\,\psi_k(x,t),\hspace{1cm}j=1,\,2,\,\dots,\,N,\label{nls1}
\end{equation}
where we have required  $(\phi_j)^T$ to be also a solution to (\ref{ncnlse2}) and ${c_j}_k$ to be arbitrary real constants. 

We consider the case of  three-coupled NLSE
\begin{eqnarray}
i\, \frac{\partial}{\partial t}\psi_1(x,t)+\frac{\partial^2}{\partial x^2}\psi_1(x,t)+\left(b_{11}\,|\psi_1(x,t)|^2+b_{12}\,|\psi_2(x,t)|^2+b_{13}\,|\psi_3(x,t)|^2\right)\,\psi_1(x,t)&=&0,\label{3cnlse1}\\\label{3cnlse2}
i\, \frac{\partial}{\partial t}\psi_2(x,t)+\frac{\partial^2}{\partial x^2}\psi_2(x,t)+\left(b_{21}\,|\psi_1(x,t)|^2+b_{22}\,|\psi_2(x,t)|^2+b_{23}\,|\psi_3(x,t)|^2\right)\,\psi_2(x,t)&=&0,\\\label{3cnlse3}
i\, \frac{\partial}{\partial t}\psi_3(x,t)+\frac{\partial^2}{\partial x^2}\psi_3(x,t)+\left(b_{31}\,|\psi_1(x,t)|^2+b_{32}\,|\psi_2(x,t)|^2+b_{33}\,|\psi_3(x,t)|^2\right)\,\psi_3(x,t)&=&0.
\end{eqnarray}
The seed exact solution of this system is denoted  by $\left(\psi_1,\psi_2, \psi_3\right)^T$. We construct  the following linear superposition of its components
\begin{eqnarray}
\phi_1(x,t)&=&c_1\,\psi_1(x,t)+c_2\,\psi_2(x,t)+c_3\,\psi_3(x,t)\label{3ls1},\\
\phi_2(x,t)&=&c_4\,\psi_1(x,t)+c_5\,\psi_2(x,t)+c_6\,\psi_3(x,t),\label{3ls2}\\
\phi_3(x,t)&=&c_7\,\psi_1(x,t)+c_8\,\psi_2(x,t)+c_9\,\psi_3(x,t),\label{3ls3}
\end{eqnarray}
where $c_{1-9}$ are arbitrary real constants. We require  $\left(\phi_1,\phi_2,\phi_3\right)^T$ to be also a solution to the system (\ref{3cnlse1})-(\ref{3cnlse3}). Substituting (\ref{3ls1})-(\ref{3ls3}) in the system (\ref{3cnlse1})-(\ref{3cnlse3}), gives
%%%%%%%%%%%%%%%%%
%%%%%%%%%%%%%%%%%%
\begin{eqnarray}
&&c_1 \left[i \psi _{1t}+\psi _{{1xx}}+\left({
	A_1|\psi _1|^2}+A_2|\psi _2|^2+A_3|\psi _3|^2
\right)\psi _1+A_4\psi_1^2\psi_2^*+A_5\psi_1^*\psi_2\psi_3+A_6\psi_1^2\psi_3^*\right]\nonumber\\
&+& c_2 \left[i \psi _{2 t}+\psi _{2
	{xx}}+\left(A_{7}|\psi _1|^2+A_{8}|\psi _2|^2+A_{9}|\psi _3|^2\right)  \psi _2+A_{10}\psi_1^*\psi_2^2+A_{11}\psi_1\psi_2^*\psi_3+A_{12}\psi_2^2\psi_3^*\right]\nonumber\\
&+& c_3 \left[i \psi _{2 t}+\psi _{2
	{xx}}+\left(A_{13}|\psi _1|^2+A_{14}|\psi _2|^2+A_{15}|\psi _3|^2\right)  \psi _3+A_{16}\psi_1^*\psi_3^2+A_{17}\psi_1\psi_2\psi_3^*+A_{18}\psi_2^*\psi_3^2\right]=0\label{3cnlse33},\nonumber\\\\\nonumber\\\nonumber\\
&&c_4 \left[i \psi _{1t}+\psi _{{1xx}}+\left({
	B_1|\psi _1|^2}+B_2|\psi _2|^2+B_3|\psi _3|^2
\right)\psi _1+B_4\psi_1^2\psi_2^*+B_5\psi_1^*\psi_2\psi_3+B_6\psi_1^2\psi_3^*\right]\nonumber\\
&+& c_5 \left[i \psi _{2 t}+\psi _{2
	{xx}}+\left(B_{7}|\psi _1|^2+B_{8}|\psi _2|^2+B_{9}|\psi _3|^2\right)  \psi _2+B_{10}\psi_1^*\psi_2^2+B_{11}\psi_1\psi_2^*\psi_3+B_{12}\psi_2^2\psi_3^*\right]\nonumber\\
&+& c_6 \left[i \psi _{2 t}+\psi _{2
	{xx}}+\left(B_{13}|\psi _1|^2+B_{14}|\psi _2|^2+B_{15}|\psi _3|^2\right)  \psi _3+B_{16}\psi_1^*\psi_3^2+B_{17}\psi_1\psi_2\psi_3^*+B_{18}\psi_2^*\psi_3^2\right]=0\label{3cnlse4},\nonumber\\\\\nonumber\\\nonumber\\
&&c_7 \left[i \psi _{1t}+\psi _{{1xx}}+\left({
	C_1|\psi _1|^2}+C_2|\psi _2|^2+C_3|\psi _3|^2
\right)\psi _1+C_4\psi_1^2\psi_2^*+C_5\psi_1^*\psi_2\psi_3+C_6\psi_1^2\psi_3^*\right]\nonumber\\
&+& c_8 \left[i \psi _{2 t}+\psi _{2
	{xx}}+\left(C_{7}|\psi _1|^2+C_{8}|\psi _2|^2+C_{9}|\psi _3|^2\right)  \psi _2+C_{10}\psi_1^*\psi_2^2+C_{11}\psi_1\psi_2^*\psi_3+C_{12}\psi_2^2\psi_3^*\right]\nonumber\\
&+& c_9 \left[i \psi _{2 t}+\psi _{2
	{xx}}+\left(C_{13}|\psi _1|^2+C_{14}|\psi _2|^2+C_{15}|\psi _3|^2\right)  \psi _3+C_{16}\psi_1^*\psi_3^2+C_{17}\psi_1\psi_2\psi_3^*+C_{18}\psi_2^*\psi_3^2\right]=0\label{3cnlse5},
\end{eqnarray}
where the coefficients $A_j$, $B_j$, and $C_j$, $j=1,2,\dots, 18$ are defined in Appendix \ref{appendixB}. 
In a similar manner, as for the two-coupled NLSE, the expressions in the square brackets of Eqs. (\ref{3cnlse33}), (\ref{3cnlse4}), and (\ref{3cnlse5}) correspond to the left hand side of Eqs. (\ref{3cnlse1}),  (\ref{3cnlse2}), and (\ref{3cnlse3}),  with extra cross terms $\psi_1^2\psi_2^*,\, \psi_1^*\psi_2\psi_3,\,\psi_1^2\psi_3^*,\,\psi_1^*\psi_2^2,\, \psi_1\psi_2^*\psi_3,\,\psi_2^2\psi_3^*,\, \psi_1^*\psi_3^2,\,\psi_1\psi_2\psi_3^*,$ and $\psi_2^*\psi_3^2$. We require the coefficients of the cross terms to vanish and the coefficients of the other nonlinear terms in Eqs. (\ref{3cnlse33}), (\ref{3cnlse4}), and (\ref{3cnlse5}) to match their counterparts in Eqs. (\ref{3cnlse1}),  (\ref{3cnlse2}), and (\ref{3cnlse3}), respectively. This will result in a set of equations for the unknown constants $c_{1-9}$ with the following solutions
\begin{eqnarray}
c_2&=&\frac{c_1 c_8\sqrt{b_{13}b_{31}}\sqrt{b_{13}(1+b_{33} c_8^2)-b_{33}c_9^2}+c_9\sqrt{b_{13} b_{31}c_1^2+b_{11} b_{13} b_{33}c_8^2-b_{11} b_{33}c_9^2}}{\sqrt{b_{11} b_{31} b_{33}}(b_{13} c_8^2-c_9^2)},\nonumber\\
c_3&=&\frac{c_1 c_9\sqrt{b_{13}b_{31}}\sqrt{b_{13}(1+b_{33} c_8^2)-b_{33}c_9^2}+c_8 b_{13}\sqrt{b_{13} b_{31}c_1^2+b_{11} b_{13} b_{33}c_8^2-b_{11} b_{33}c_9^2}}{\sqrt{b_{11} b_{31} b_{33}}(b_{13} c_8^2-c_9^2)},\nonumber\\
c_4&=&\frac{\sqrt{b_{13} b_{31}c_1^2+b_{11} b_{13} b_{33}c_8^2-b_{11} b_{33}c_9^2}}{\sqrt{b_{13}}},\nonumber\\
c_5&=&\frac{c_1 c_9\sqrt{b_{13}b_{31}}+c_8\sqrt{b_{13}(1+b_{33} c_8^2)-b_{33}c_9^2}\sqrt{b_{13} b_{31}c_1^2+b_{11} b_{13} b_{33}c_8^2-b_{11} b_{33}c_9^2}}{\sqrt{b_{11} b_{33}}(b_{13} c_8^2-c_9^2)},\nonumber\\
c_6&=&\frac{c_1 c_8 b_{13}^{3/2}\sqrt{b_{31}}+c_9\sqrt{b_{13}(1+b_{33} c_8^2)-b_{33}c_9^2}\sqrt{b_{13} b_{31}c_1^2+b_{11} b_{13} b_{33}c_8^2-b_{11} b_{33}c_9^2}}{\sqrt{b_{11} b_{33}}(b_{13} c_8^2-c_9^2)},\nonumber\\
c_7&=&\frac{\sqrt{b_{11}}\sqrt{b_{13}(1+b_{33} c_8^2)-b_{33}c_9^2}}{\sqrt{b_{13} b_{33}}},
\end{eqnarray}
\begin{equation}
b_{12}=b_{22}=b_{32}=-1,
\end{equation}
where $c_1$, $c_8$, and $c_9$ remain to be arbitrary real constants. Substituting these coefficients back into Eqs. (\ref{3ls1}), (\ref{3ls2}), and (\ref{3ls3}), we obtain the composite solution 
\begin{eqnarray}
\phi_1(x,t)&=&c_1\,\psi_1(x,t)+\frac{c_1 c_8\sqrt{b_{13}b_{31}}\sqrt{b_{13}(1+b_{33} c_8^2)-b_{33}c_9^2}+c_9\sqrt{b_{13} b_{31}c_1^2+b_{11} b_{13} b_{33}c_8^2-b_{11} b_{33}c_9^2}}{\sqrt{b_{11} b_{31} b_{33}}(b_{13} c_8^2-c_9^2)}\,\psi_2(x,t)\nonumber\\&&+\frac{c_1 c_9\sqrt{b_{13}b_{31}}\sqrt{b_{13}(1+b_{33} c_8^2)-b_{33}c_9^2}+c_8 b_{13}\sqrt{b_{13} b_{31}c_1^2+b_{11} b_{13} b_{33}c_8^2-b_{11} b_{33}c_9^2}}{\sqrt{b_{11} b_{31} b_{33}}(b_{13} c_8^2-c_9^2)}\,\psi_3(x,t)\label{3ls11},\\
\phi_2(x,t)&=&\frac{\sqrt{b_{13} b_{31}c_1^2+b_{11} b_{13} b_{33}c_8^2-b_{11} b_{33}c_9^2}}{\sqrt{b_{13}}}\,\psi_1(x,t)\nonumber\\&&+\frac{c_1 c_9\sqrt{b_{13}b_{31}}+c_8\sqrt{b_{13}(1+b_{33} c_8^2)-b_{33}c_9^2}\sqrt{b_{13} b_{31}c_1^2+b_{11} b_{13} b_{33}c_8^2-b_{11} b_{33}c_9^2}}{\sqrt{b_{11} b_{33}}(b_{13} c_8^2-c_9^2)}\,\psi_2(x,t)\nonumber\\&&+\frac{c_1 c_8 b_{13}^{3/2}\sqrt{b_{31}}+c_9\sqrt{b_{13}(1+b_{33} c_8^2)-b_{33}c_9^2}\sqrt{b_{13} b_{31}c_1^2+b_{11} b_{13} b_{33}c_8^2-b_{11} b_{33}c_9^2}}{\sqrt{b_{11} b_{33}}(b_{13} c_8^2-c_9^2)}\,\psi_3(x,t),\label{3ls22}\\
\phi_3(x,t)&=&\frac{\sqrt{b_{11}}\sqrt{b_{13}(1+b_{33} c_8^2)-b_{33}c_9^2}}{\sqrt{b_{13} b_{33}}}\,\psi_1(x,t)+c_8\,\psi_2(x,t)+c_9\,\psi_3(x,t),\label{3ls33}
\end{eqnarray}
which satisfy the three-coupled NLSE
\begin{eqnarray}
i\, \frac{\partial}{\partial t}\phi_1(x,t)+\frac{\partial^2}{\partial x^2}\phi_1(x,t)+\left(b_{11}\,|\phi_1(x,t)|^2+b_{12}\,|\phi_2(x,t)|^2+b_{13}\,|\phi_3(x,t)|^2\right)\,\phi_1(x,t)&=&0,\label{3cnlse111}\\\label{3cnlse222}
i\, \frac{\partial}{\partial t}\phi_2(x,t)+\frac{\partial^2}{\partial x^2}\phi_2(x,t)+\left(b_{21}\,|\phi_1(x,t)|^2+b_{22}\,|\phi_2(x,t)|^2+b_{23}\,|\phi_3(x,t)|^2\right)\,\phi_2(x,t)&=&0,\\\label{3cnlse333}
i\, \frac{\partial}{\partial t}\phi_3(x,t)+\frac{\partial^2}{\partial x^2}\phi_3(x,t)+\left(b_{31}\,|\phi_1(x,t)|^2+b_{32}\,|\phi_2(x,t)|^2+b_{33}\,|\phi_3(x,t)|^2\right)\,\phi_3(x,t)&=&0.
\end{eqnarray}
Similar to the previous case of two-coupled NLSE, the conservation of the total norm squared is guaranteed by the matching of nonlinear coefficients in Eqs. (\ref{3cnlse1})-(\ref{3cnlse3}) to those of Eqs. (\ref{3cnlse111})-(\ref{3cnlse333}), which results in
\begin{equation}
b_{11}|\phi_1|^2+b_{12}|\phi_2|^2+b_{13}|\phi_3|^2=b_{11}|\psi_1|^2+b_{12}|\psi_2|^2+b_{13}|\psi_3|^2.
\end{equation}
This conservation law introduces a rotation symmetry. Similar to the previous case of two-coupled NLSE, the composite solution can be represented in terms of a rotation in a three-dimensional function space. One needs in this case to rewrite the remaining free constants, $c_1,\,c_8,\,c_9$, in terms of rotation angles. However, there is no need for such representation since, as we will see in the next section, all seed solutions we use in this case are obtained by mapping the three-coupled NLSE to two-coupled NLSE where one of the components, say $\psi_2$, is proportional to another one, say $\psi_2=\sigma\,\psi_1$ with $\sigma$ being arbitrary complex constant. As a result, the three-dimensional function space reduces to two-dimensional function space and hence the rotation representation in the two cases will be essentially the same. The same arguments apply for the $N$-coupled NLSE.

\subsection{Symmetry reduction}
The three-coupled NLSE, Eqs. (\ref{3cnlse1})-(\ref{3cnlse3}), can be reduced to the followoing two-coupled NLSE
\begin{eqnarray}
i\, \frac{\partial}{\partial t}\psi_1(x,t)+\frac{\partial^2}{\partial x^2}\psi_1(x,t)+\left(b_{11M}\,|\psi_1(x,t)|^2+b_{12M}\,|\psi_3(x,t)|^2\right)\,\psi_1(x,t)&=&0,\label{cnlse111}\\\label{cnlse222}
i\, \frac{\partial}{\partial t}\psi_3(x,t)+\frac{\partial^2}{\partial x^2}\psi_3(x,t)+\left(b_{21M}\,|\psi_1(x,t)|^2+b_{22M}\,|\psi_3(x,t)|^2\right)\,\psi_3(x,t)&=&0,
\end{eqnarray}
with the following replacements \cite{ourbook}
\begin{enumerate}
	\item{$\psi_2(x,t)=\sigma \psi_1(x,t)$,}
	\item{$b_{11}={b_{11}}_{M}-b_{12}|\sigma|^2$,}
	\item{$b_{21}=b_{11}+(b_{12}-b_{22})|\sigma|^2$,}
	\item{$b_{23}=b_{13}={b_{12}}_{M}$,}
	\item{$b_{31}={b_{21}}_{M}-b_{32}|\sigma|^2$,}
	\item{$b_{33}={b_{22}}_{M}$,}
\end{enumerate}
where $\sigma$ is an  arbitrary complex constant and ${b_{11}}_{M}$, ${b_{12}}_{M}$, ${b_{21}}_{M}$, and ${b_{22}}_{M}$ are the constant coefficients of the nonlinear terms in the Manakov system. Such a symmetry reduction establishes a one-to-one correspondence  between the solutions of the two-coupled NLSE and  those of the three-coupled NLSE. In the following, we employ this symmetry reduction to obtain composite solutions of the three-coupled NLSE using either the seed solution or the composite solution of Solution 4 of the two-coupled NLSE. Therefore, we consider two examples. In Example 1, we first apply the above symmetry reduction to the seed solution of the two-coupled NLSE, (\ref{seed41}) and  (\ref{seed42}), and then use the result, $(\psi_1,\psi_2, \psi_3)^T$, as a seed solution to  the transformation (\ref{3ls11})-(\ref{3ls33}) leading to the corresponding composite solution $(\phi_1,\phi_2, \phi_3)^T$.  In Example 2, we apply the symmetry reduction to the composite solution, (\ref{comp41}) and (\ref{comp42}), and
then use the result, $(\psi_1,\psi_2,\psi_3)^T$, as a seed solution to the transformation (75)-(77) leading to the corresponding
composite solution $(\phi_1,\phi_2,\phi_3)^T$.\\

\begin{widetext}
\begin{figure}[!h]
	\centering
	\includegraphics[scale=1]{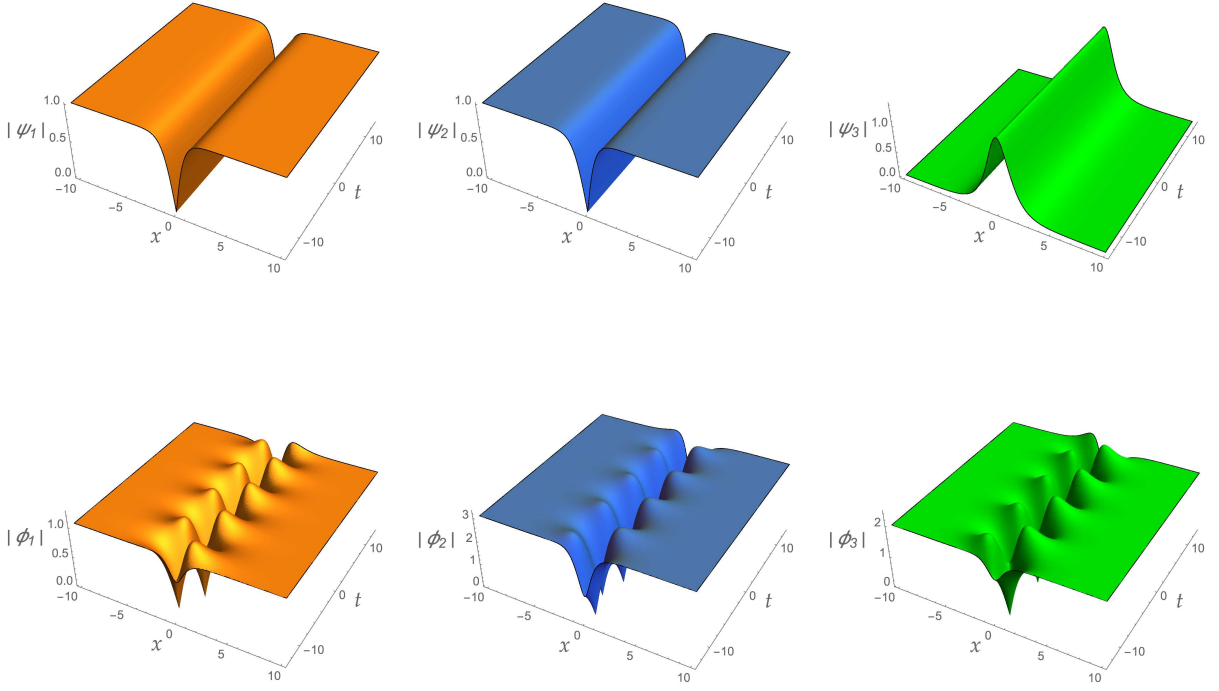}
	\caption{The norm of seed (\ref{3seed41})-(\ref{3seed42})  (upper pannel) and its composite solution given by Eqs. (\ref{3seed41comp})-(\ref{3seed42comp}) (lower pannel). Orange corresponds to $\psi_1$ and $\phi_1$, blue corresponds to $\psi_2$ and $\phi_2$, and green corresponds to $\psi_3$ and $\phi_3$. Parameters used: $A_0=c_1=c_8=c_9=1$ and $\sigma=i$.}
	\label{fig10}
\end{figure}
\end{widetext}

\begin{figure}[!h]
	\centering
	\includegraphics[scale=1]{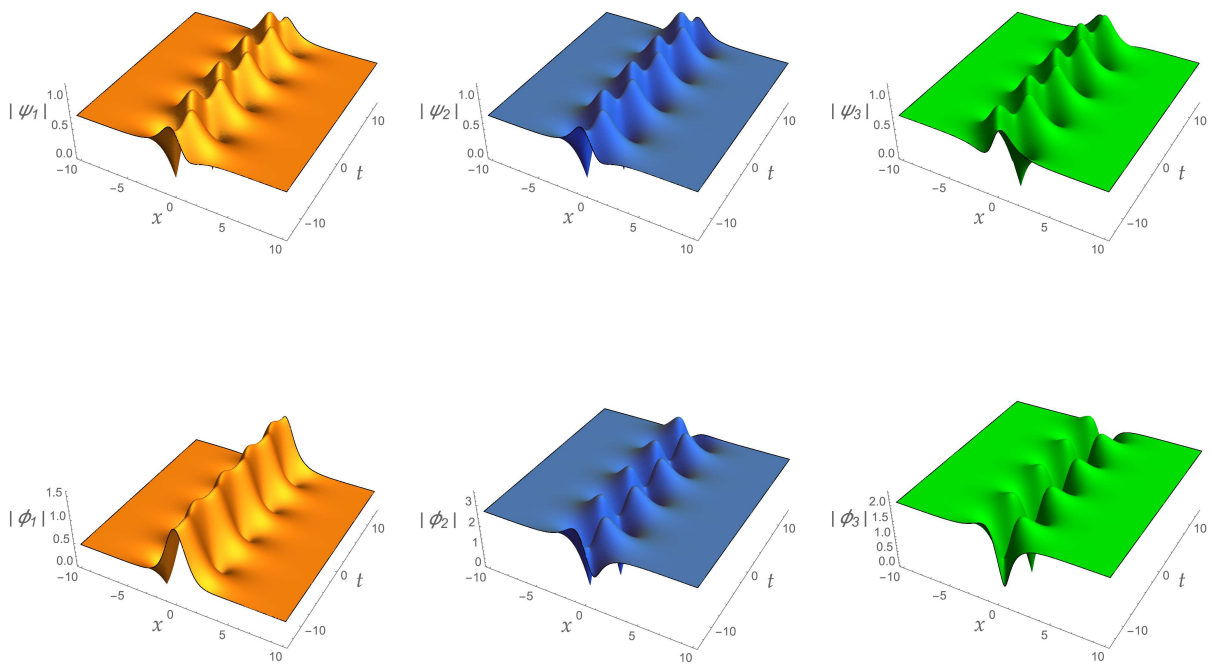}
	\caption{The norm of seed (\ref{3comp41})-(\ref{3comp42})  (upper pannel) and its composite solution  given by Eqs. (\ref{32comp41})-(\ref{32comp42}) (lower pannel). Orange corresponds to $\psi_1$ and $\phi_1$, blue corresponds to $\psi_2$ and $\phi_2$, and green corresponds to $\psi_3$ and $\phi_3$. Parameters used: $A_0=c_1=c_8=c_9=1$, $\sigma=i$, and $\theta=\pi/4$ for the composite solution.}
	\label{fig11}
\end{figure}

\textit{Example 1. Dark-dark-bright soliton:} This solution can be constructed from the seed solution in (\ref{seed41}) and  (\ref{seed42}). It  corresponds  to  three-coupled stationary solitons; two dark solitons and a single bright soliton.  In Fig. \ref{fig10}, we show the seed solution components together with their composite solution components, where the stationary dark and the bright solitons turn into  breathers. The seed solution reads in this case
\begin{equation}
\psi_1=A_0\, e^{2iA_0^2t}{\rm tanh}(A_0 x)\label{3seed41},
\end{equation}
\begin{equation}
\psi_2=\sigma\psi_1\label{33seed41},
\end{equation}
\begin{equation}
\psi_3=-i\sqrt{2}\,A_0\, e^{3iA_0^2t}{\rm sech}(A_0 x)\label{3seed42},
\end{equation}
where $A_0$ is an arbitrary real constant and ${b_{11}}_{M}={b_{12}}_{M}={b_{21}}_{M}={b_{22}}_{M}=2$. The transformation (\ref{3ls11})-(\ref{3ls33}) yields its composite solution as 
\begin{eqnarray}
\phi_1&=&\frac{A_0}{D_2}\Big[-2i\,D_0\,e^{iA_0^2t}+D_1\,\text{sinh}(A_0 x)\Big]\text{sech}(A_0 x)e^{2iA_0^2t}\label{3seed41comp},
\end{eqnarray}
\begin{eqnarray}
\phi_2&=&\frac{A_0}{D_5}\Big[-2i\,D_3\,e^{iA_0^2t}+D_4\,\text{sinh}(A_0 x)\Big]\text{sech}(A_0 x)e^{2iA_0^2t}\label{33seed41comp},
\end{eqnarray}
\begin{equation}
\phi_3=\frac{A_0}{2}\left[-2i\,\sqrt{2}c_9\,e^{iA_0^2t}+D_6\,\text{sinh}(A_0 x)\right]\text{sech}(A_0 x)e^{2iA_0^2t}\label{3seed42comp},
\end{equation}
where\\\\
$\begin{aligned}
 D_0={}&c_1c_9\sqrt{1+2c_8^2-c_9^2}\sqrt{2+|\sigma|^2}\\&+c_8\sqrt{2(c_1^2+2c_8^2-c_9^2)(2+|\sigma|^2)},
 \end{aligned}$\\
 $\begin{aligned}
  D_1={}&c_9\sigma\sqrt{(c_1^2+2c_8^2-c_9^2)(2+|\sigma|^2)}\\&+c_1\Big[c_8\sigma\sqrt{(2+4c_8^2-2c_9^2)(2+|\sigma|^2)}+(2c_8^2-c_9^2)(2+|\sigma|^2)\Big],
  \end{aligned}$\\
 $\begin{aligned}
D_2={}&(2c_8^2-c_9^2)(2+|\sigma|^2),  
\end{aligned}$\\
 $\begin{aligned}
D_3={}&c_1c_8\sqrt{2(2+|\sigma|^2)}\\&+c_9\sqrt{(1+2c_8^2-c_9^2)(c_1^2+2c_8^2-c_9^2)(2+|\sigma|^2)},
\end{aligned}$\\
 $\begin{aligned}
D_4={}&c_1c_9\sigma\sqrt{2+|\sigma|^2}\\&+\sqrt{(c_1^2+2c_8^2-c_9^2)(2+|\sigma|^2)}\\&\times\Big(c_8\sigma\sqrt{2+4c_8^2-2c_9^2}+2c_8^2\sqrt{2+|\sigma|^2}-c_9^2\sqrt{2+|\sigma|^2}\Big),
\end{aligned}$\\
 $\begin{aligned}
D_5={}&(2c_8^2-c_9^2)\sqrt{2+|\sigma|^2},
\end{aligned}$
\\\\
and\\\\
$\begin{aligned}
D_6={}&2c_8\sigma+\sqrt{(2+4c_8^2-2c_9^2)(2+|\sigma|^2)}.
\end{aligned}$\\\\
\textit{Example 2. Bright-bright-dark soliton:} This solution can be constructed from the composite solution in Eqs.  (\ref{comp41}) and (\ref{comp42}). Figure \ref{fig11} shows that the three components of both the seed and the composite solution  for $\theta=\pi/4$ are breathers. The seed solution takes the form
\begin{equation}
\psi_1=A_0\, e^{2iA_0^2t}\left[{\rm tanh}(A_0 x)\,\cos\theta-i\sqrt{2}\, e^{iA_0^2t}{\rm sech}(A_0 x)\,\sin\theta\right]\label{3comp41},
\end{equation}
\begin{equation}
\psi_2=\sigma\psi_1\label{33comp41},
\end{equation}
\begin{equation}
\psi_3=-A_0\, e^{2iA_0^2t}\left[{\rm tanh}(A_0 x)\,\sin\theta+i\sqrt{2}\, e^{iA_0^2t}{\rm sech}(A_0 x)\,\cos\theta\right]\label{3comp42},
\end{equation}
where $A_0$ is an arbitrary real constant and ${b_{11}}_{M}={b_{12}}_{M}={b_{21}}_{M}={b_{22}}_{M}=2$. The transformation (\ref{3ls11})-(\ref{3ls33}) leads to the following composite solution
\begin{widetext}
\begin{eqnarray}
\phi_1&=&\frac{A_0e^{2iA_0^2t}}{D_{9}} \Big[D_7\left(\sqrt{2}e^{iA_0^2t}\text{sech}(A_0 x)\text{sin}(\theta)+i\text{tanh}(A_0 x)\text{cos}(\theta)\right)\\\nonumber&&-D_{8}\,\left(\sqrt{2}e^{iA_0^2t}\text{sech}(A_0 x)\text{cos}(\theta)-i\text{tanh}(A_0 x)\text{sin}(\theta)\right)\Big]\label{32comp41},
\end{eqnarray}
\begin{eqnarray}
\phi_2&=&\frac{A_0e^{2iA_0^2t}}{D_{12}}\Big[D_{10}\,\left(\sqrt{2}e^{iA_0^2t}\text{sech}(A_0 x)\text{sin}(\theta)+i\text{tanh}(A_0 x)\text{cos}(\theta)\right)\nonumber\\&&-D_{11}\,\left(\sqrt{2}e^{iA_0^2t}\text{sech}(A_0 x)\text{cos}(\theta)-i\text{tanh}(A_0 x)\text{sin}(\theta)\right)\Big]\label{332comp41},
\end{eqnarray}
\begin{eqnarray}
\phi_3&=&\frac{A_0e^{2iA_0^2t}}{2}\Bigg[\text{tanh}(A_0 x)\Big(D_{13}\,\text{cos}(\theta)-2c_9\,\text{sin}(\theta)\Big)-2ie^{iA_0^2t}\text{sech}(A_0 x)\Big(\sqrt{2}c_9\,\text{cos}(\theta)+D_{14}\,\text{sin}(\theta)\Big)\Bigg]\label{32comp42},
\end{eqnarray}
where\\\\
$\begin{aligned}
D_7={}&-i\sigma\sqrt{2}\Big[2c_1c_8\sqrt{(1+2c_8^2-c_9^2)(2+|\sigma|^2)}+c_9\sqrt{2(c_1^2+2c_8^2-c_9^2)(2+|\sigma|^2)}\Big]-2ic_1(2c_8^2-c_9^2)(2+|\sigma|^2),
\end{aligned}$\\
$\begin{aligned}
D_8={}&2i\sqrt{2}\Big[c_1c_9\sqrt{(1+2c_8^2-c_9^2)(2+|\sigma|^2)}+c_8\sqrt{2(c_1^2+2c_8^2-c_9)(2+|\sigma|^2)}\Big],
\end{aligned}$\\
$\begin{aligned}
D_9={}&2(2c_8^2-c_9^2)(2+|\sigma|^2),
\end{aligned}$\\
$\begin{aligned}
D_{10}={}&-i\sigma\Big[c_1c_9\sqrt{2(2+|\sigma|^2)}+2c_8\sqrt{(1+2c_8^2-c_9^2)(c_1^2+2c_8^2-c_9^2)(2+|\sigma|^2)}\Big]-i(2c_8^2-c_9^2)(2+|\sigma|^2)\sqrt{2(c_1^2+2c_8^2-c_9^2)},
\end{aligned}$\\
$\begin{aligned}
D_{11}={}&2i\Big[c_1c_8\sqrt{2(2+|\sigma|^2)}+c_9\sqrt{(1+2c_8^2-c_9^2)(c_1^2+2c_8^2-c_9^2)(2+|\sigma|^2)}\Big],
\end{aligned}$\\
$\begin{aligned}
D_{12}={}&(2c_8^2-c_9^2)\sqrt{2(2+|\sigma|^2)},  
\end{aligned}$\\
$\begin{aligned}
D_{13}={}&2c_8\sigma+\sqrt{(2+4c_8^2-2c_9^2)(2+|\sigma|^2)},
\end{aligned}$\\\\
and\\\\
$\begin{aligned}
D_{14}={}&\sqrt{2}c_8\sigma+\sqrt{(1+2c_8^2-c_9^2)(2+|\sigma|^2)}.
\end{aligned}$\\\\
\end{widetext}

\section{Conclusions and Outlook}
\label{concsec}
We have considered an $N$-coupled NLSE with emphasis on $N=2$ and $N=3$. A solution of an $N$-coupled NLSE is given in the form of a vector with $N$ components $\psi_i(x,t),\,\,i=1,2,\dots,N$.  Given a vector solution, which we denote as a seed solution, a linear combination of its components, which we denote as a composite solution, may also be a solution to the same coupled NLSE under some restrictions on the constant coefficients of the linear combination. Taking the resulting composite solution as a seed solution and constructing a linear combination of its components generates another composite solution. Repeated action of this procedure leads to a series of composite solutions, which we denote as a family of solutions. Using another seed generates another family such that each family is characterized by its seed solution. The number of solutions in each family could be finite or infinite. Due to the fact that composite solutions preserve the total square norm, $\sum_{i=1}^N|\psi_i|^2$, a rotational symmetry exists among the composite solutions of a given family. It turns out that all members of a family of solutions can be obtained by repeated action of a rotation operator on the seed solution. The angle of rotation, $\theta$, is arbitrary. If the angle is in the form of $\theta=2\pi/n$, where $n$ is an integer, the number of composite solutions in the family will be equal to $n$ because after $n$ applications of the rotation operator, the solutions start to repeat. If $n$ is not an integer, the number of solutions will be infinite. 

Then, we have considering two composite solutions that belong to the same family and are obtained by $\theta_1$ and $\theta_2$ rotations.  A superposition of these two composite solutions was shown to be also a solution under certain restrictions on the coefficients. It turned out that such superposition can be also obtained by applying the rotation operator on the seed solution with an angle $\delta$ that depends on both $\theta_1$ and $\theta_2$. This means that the superposition of two composite solutions is another composite solution of the same family and hence we conclude that the superposition is equivalent to the application of the rotation operator. 

For the specific case of two-coupled Manakov system ($N=2$), many exact solutions are known. We have considered almost all of these solutions as seed solutions and generated for each one its family of composite solutions. For most cases, the generated composite solutions possessed features that do not exist in the seed solutions. In general, if the components of the seed solution are stationary localized functions in a uniform background, the composite solution turns to either a breather in a uniform background or a breather in an oscillatory background. Here we point out the main features for each case considered. In Solution 1, the two components of the seed solution are CW. Remarkably, the two components of its composite solution are out-of-phase traveling waves, as shown in Fig. \ref{fig2}. Similar behavior is obtained in Solution 2 where traveling waves are introduced to the decaying background, as shown in Fig. \ref{fig3}. Interesting result is obtained in Solution 3 where the seed solution representing the scattering of a dark and bright solitons with a Peregrine soliton in a uniform background, turn to a scattering of two out-of-phase breathers with the Peregrine soliton, as shown in Fig. \ref{fig4}. Solution 4, represents a special limiting case of Solution 3, which corresponds to a dark-bright soliton solution without the Peregrine soliton, as shown in Fig. \ref{fig5}. Solutions 5 and 6 correspond to higher order Peregrine solitons. While the components of the seed solution are higher order Peregrine solitons with a uniform background, the components of the composite solution are higher order Peregrine solitons with a background of two out-of-phase traveling waves, as shown in Figs. \ref{fig6} and \ref{fig7}.  Solution 7 corresponds to two bright solitons where, similar to Solution 4, the stationary seed solitons turn to breathers, as shown in Fig. \ref{fig8}. Solution 8 corresponds to a seed solution with two components that are stationary solitary waves represented by Jacobi elliptic functions. The corresponding composite solution turns out to be a breather that is oscillating in both space and time, as shown in Fig. \ref{fig9}. It is pointed out that this kind of breather is complementary to the Akhmediev \cite{akhm}, Kusnetz-Ma \cite{km}, and Peregrine \cite{per} breathers. 

Finally, we have considered coupled NLSE with $N>2$. For the case of general $N$, one can use some symmetry reductions to reduce the system to the two-coupled NLSE. Therefore, all composite solutions obtained above for the two-coupled NLSE can be mapped to the $N$-coupled NLSE. We took a specific case of $N=3$ and have shown this to be true by deriving the composite solution for an example of three solitons. Similar to the case of two-coupled NLSE, the stationary localized solution turned to a breather.

In addition to the interesting features brought by the composite solutions of the coupled NLSE, we believe the superposition principle may be helpful in the attempts to find a series expansion of  the solution in terms of some orthogonal functions. Another possible application would be the NLSE with dual power nonlinearity where nonlinear cross terms may also cancel each other allowing for the application of the superposition principle on that case as well.

\section*{Acknowledgments} We acknowledge the support of UAE University through grants UAEU-UPAR(4) 2016 and UAEU-UPAR(6) 2017.

\appendix 

\section{Seed solution 6 of two-coupled NLSE}
\label{appendixA}
\begin{eqnarray}
	\psi_1&=&A_0\,\Big[\frac{\alpha_1(x,t)+i\,\beta_1(x,t)}{\gamma(x,t)}-(1+i\,\sqrt{3})\Big]\nonumber\\&&\times e^{i\,[A_1\,x+(16\,A_0^2-A_1^2)\,t]},\\
	\psi_2&=&A_0\,\Big[\frac{\alpha_2(x,t)+i\,\beta_2(x,t)}{\gamma(x,t)}-(1-i\,\sqrt{3})\Big]\nonumber\\&&\times e^{i\,[A_2\,x+(16\,A_0^2-A_2^2)\,t]},
\end{eqnarray}
\begin{widetext}
where \\\\
$\begin{aligned}
\alpha_1(x,t)={}&-864\,\sqrt{3}\,A_0^6\,t^3-144\,\sqrt{3}\,A_0^5\,\delta(x,t)\,t^2-72\,\sqrt{3}\,A_0^4\,\delta^2(x,t)\,t-216\,A_0^4\,t^2\\&-12\,\sqrt{3}\,A_0^3\,\delta^3(x,t)-144\,A_0^3\,\delta(x,t)\,t-18\,A_0^2\,\delta^2(x,t)-12\,\sqrt{3}\,A_0^2\,t+3,
\end{aligned}$\\
\\\\
$\begin{aligned}
\alpha_2(x,t)={}&864\,A_0^6\,\sqrt{3}\,t^3-144\,\sqrt{3}\,A_0^5\,\delta(x,t)\,t^2+72\,\sqrt{3}\,A_0^4\,\delta^2(x,t)\,t-216\,A_0^4\,t^2\\&-12\,\sqrt{3}\,A_0^3\,\delta^3(x,t)+144\,A_0^3\,\delta(x,t)\,t-18\,A_0^2\,\delta^2(x,t)+12\,\sqrt{3}\,A_0^2\,t+3,
\end{aligned}$\\
\\\\
$\begin{aligned}
\beta_1(x,t)={}&864\,A_0^6\,t^3+144\,A_0^5\,\delta(x,t)\,t^2+72\,A_0^4\,\delta^2(x,t)\,t+312\,\sqrt{3}\,A_0^4\,t^2+12\,A_0^3\,\delta^3(x,t)\\&+96\,\sqrt{3}\,A_0^3\,\delta(x,t)\,t+18\,\sqrt{3}\,A_0^2\,\delta^2(x,t)+108\,A_0^2\,t+12\,A_0\,\delta(x,t)+\sqrt{3},
\end{aligned}$\\
\\\\
$\begin{aligned}
\beta_2(x,t)={}&864\,A_0^6\,t^3-144\,A_0^5\,\delta(x,t)\,t^2+72\,A_0^4\,\delta^2(x,t)\,t-312\,\sqrt{3}\,A_0^4\,t^2-12\,A_0^3\,\delta^3(x,t)\\&+96\,\sqrt{3}\,A_0^3\,\delta(x,t)\,t-18\,\sqrt{3}\,A_0^2\,\delta^2(x,t)+108\,A_0^2\,t-12\,A_0\,\delta(x,t)-\sqrt{3},
\end{aligned}$\\
\\\\
$\begin{aligned}
\gamma(x,t)={}&1728\,A_0^8\,t^4+288\,A_0^6\,\delta^2(x,t)\,t^2+384\,\sqrt{3}\,A_0^5\,\delta(x,t)\,t^2+12\,A_0^4\,\delta^4(x,t)+432\,A_0^4\,t^2\\&+16\,\sqrt{3}\,A_0^3\,\delta^3(x,t)+24\,A_0^2\,\delta^2(x,t)+4\,\sqrt{3}\,A_0\,\delta(x,t)+1,
\end{aligned}$\\\\\\
$A_0=A_2+3\,A_3$, $A_1=A_2-2\,A_0$, $A_2=-6A_3$,  $\delta(x,t)=x+6\,A_3\,t$, and $A_3$ is an arbitrary real constant.
\end{widetext}
\section{Coefficients of superposition transformation in the three-coupled NLSE}
 \label{appendixB}

\begin{eqnarray}
A_1&=&b_{11} c_1^2+b_{12} c_4^2 + b_{13} c_7^2,\nonumber\\A_2&=&\frac{2b_{11} c_1 c_2^2+b_{12} c_5(c_2 c_4+c_1 c_5)+b_{13} c_8 (c_2 c_7+c_1 c_8)}{c_1},\nonumber\\
A_3&=&\frac{2b_{11} c_1 c_3^2+b_{12} c_6(c_3 c_4+c_1 c_6)+b_{13} c_9 (c_3 c_7+c_1 c_9)}{c_1}
,\nonumber\\
A_4&=&\,b_{11} c_1 c_2+b_{12} c_4 c_5+b_{13} c_7 c_8,\nonumber\\
A_5&=&\frac{2b_{11} c_1 c_3 c_2+b_{12} c_4(c_3 c_5+c_2 c_6)+b_{13} c_7 (c_3 c_8+c_2 c_9)}{c_1}
,\nonumber\\
A_6&=&\,b_{11} c_1 c_3+b_{12} c_4 c_6+b_{13} c_7 c_9,\nonumber\\
A_7&=&\frac{2b_{11} c_1^2 c_2+b_{12} c_4(c_2 c_4+c_1 c_5)+b_{13} c_7 (c_2 c_7+c_1 c_8)}{c_2}
,\nonumber\\
A_8&=&b_{11} c_2^2+b_{12} c_5^2+b_{13} c_8^2,\nonumber\\
A_9&=&\frac{2b_{11} c_2 c_3^2+b_{12} c_6(c_3 c_5+c_2 c_6)+b_{13} c_9 (c_3 c_8+c_2 c_9)}{c_2},\nonumber\\
A_{10}&=&A_{4},\nonumber\\
A_{11}&=&\frac{2b_{11} c_1 c_2 c_3+b_{12} c_5(c_3 c_4+c_1 c_6)+b_{13} c_8 (c_3 c_7+c_1 c_9)}{c_2},\nonumber\\
A_{12}&=&\,b_{11} c_2 c_3+b_{12} c_5 c_6+b_{13} c_8 c_9,\nonumber\\
A_{13}&=&\frac{2b_{11} c_1^2 c_3+b_{12} c_4(c_3 c_4+c_1 c_6)+b_{13} c_7 (c_3 c_7+c_1 c_9)}{c_3},\nonumber\\
A_{14}&=&\frac{2b_{11} c_2^2 c_3+b_{12} c_5(c_3 c_5+c_2 c_6)+b_{13} c_8 (c_3 c_8+c_2 c_9)}{c_3},\nonumber\\
A_{15}&=&b_{11} c_3^2+b_{12} c_6^2+b_{13} c_9^2,\nonumber\\
A_{16}&=&A_{6},\nonumber\\
A_{17}&=&\frac{2b_{11} c_1 c_2 c_3+b_{12} c_6(c_2 c_4+c_1 c_5)+b_{13} c_9 (c_2 c_7+c_1 c_8)}{c_3},\nonumber\\
A_{18}&=&A_{12},
\nonumber\\
%%%%%%%%%%%%%%%%%%%%%%%%%%%%%B%%%%%%%%%%%%%%%%%%%%%%
B_1&=&b_{21} c_1^2+b_{22} c_4^2 + b_{23} c_7^2,
\nonumber\\
B_2&=&\frac{2b_{22} c_4 c_5^2+b_{21} c_2(c_2 c_4+c_1 c_5)+b_{23} c_8 (c_5 c_7+c_4 c_8)}{c_4},\nonumber\\
B_3&=&\frac{2b_{22} c_4 c_6^2+b_{21} c_3(c_3 c_4+c_1 c_6)+b_{23} c_9 (c_6 c_7+c_4 c_9)}{c_4},
\nonumber\\
B_4&=&b_{21} c_1 c_2+b_{22} c_4 c_5+b_{23} c_7 c_8,\nonumber\\
B_5&=&\frac{2b_{22} c_4 c_5 c_6+b_{21} c_1(c_3 c_5+c_2 c_6)+b_{23} c_7 (c_6 c_8+c_5 c_9)}{c_4},
\nonumber\\
B_6&=&b_{21} c_1 c_3+b_{22} c_4 c_6+b_{23} c_7 c_9,\nonumber\\
B_7&=&\frac{2b_{22} c_4^2 c_5+b_{21} c_1(c_2 c_4+c_1 c_5)+b_{23} c_7 (c_5 c_7+c_4 c_8)}{c_5},
\nonumber\\
B_8&=&b_{21} c_2^2+b_{22} c_5^2+b_{23} c_8^2,\nonumber\\
B_9&=&\frac{2b_{22} c_5 c_6^2+b_{21} c_3(c_3 c_5+c_2 c_6)+b_{23} c_9 (c_6 c_8+c_5 c_9)}{c_5},
\nonumber\\
B_{10}&=&B_{4},\nonumber\\
B_{11}&=&\frac{2b_{22} c_4 c_5 c_6+b_{21} c_2(c_3 c_4+c_1 c_6)+b_{23} c_8 (c_6 c_7+c_4 c_9)}{c_5},
\nonumber\\
B_{12}&=&b_{21} c_2 c_3+b_{22} c_5 c_6+b_{23} c_8 c_9,\nonumber\\
B_{13}&=&\frac{2b_{22} c_4^2 c_6+b_{21} c_1(c_3 c_4+c_1 c_6)+b_{23} c_7 (c_6 c_7+c_4 c_9)}{c_6},
\nonumber\\
B_{14}&=&\frac{2b_{22} c_5^2 c_6+b_{21} c_2(c_3 c_5+c_2 c_6)+b_{23} c_8 (c_6 c_8+c_5 c_9)}{c_6},
\nonumber\\
B_{15}&=&b_{21} c_3^2+b_{22} c_6^2+b_{23} c_9^2,\nonumber\\
A_{16}&=&A_{6},\nonumber\\
B_{17}&=&\frac{2b_{22} c_4 c_5 c_6+b_{21} c_3(c_2 c_4+c_1 c_5)+b_{23} c_9 (c_5 c_7+c_4 c_8)}{c_6},\nonumber\\B_{18}&=&B_{12},
\nonumber\\
%%%%%%%%%%%%%%%%%%%%%%%%%%%C%%%%%%%%%%%%%%%%%%%%%%%%%
C_1&=&b_{31} c_1^2+b_{32} c_4^2 + b_{33} c_7^2,\nonumber\\
C_2&=&\frac{2b_{33} c_7 c_8^2+b_{32} c_5(c_5 c_7+c_4 c_8)+b_{31} c_2 (c_2 c_7+c_1 c_8)}{c_7},\nonumber\\
C_3&=&\frac{2b_{33} c_7 c_9^2+b_{32} c_6(c_6 c_7+c_4 c_9)+b_{31} c_3 (c_3 c_7+c_1 c_9)}{c_7},\nonumber\\
C_4&=&b_{31} c_1 c_2+b_{32} c_4 c_5+b_{33} c_7 c_8,\nonumber\\
C_5&=&\frac{2b_{33} c_7 c_8 c_9+b_{32} c_4(c_6 c_8+c_5 c_9)+b_{31} c_1 (c_3 c_8+c_2 c_9)}{c_7},\nonumber\\
C_6&=&b_{31} c_1 c_3+b_{32} c_4 c_6+b_{33} c_7 c_9,\nonumber\\
C_7&=&\frac{2b_{33} c_7^2 c_8+b_{31} c_1(c_2 c_7+c_1 c_8)+b_{32} c_4 (c_5 c_7+c_4 c_8)}{c_8},\nonumber\\
C_8&=&b_{31} c_2^2+b_{32} c_5^2+b_{33} c_8^2,\nonumber\\
C_9&=&\frac{2b_{33} c_8 c_9^2+b_{31} c_3(c_3 c_8+c_2 c_9)+b_{32} c_6 (c_6 c_8+c_5 c_9)}{c_8},\nonumber\\
C_{10}&=&C_{4},\nonumber\\
C_{11}&=&\frac{2b_{33} c_7 c_8 c_9+b_{31} c_2(c_3 c_7+c_1 c_9)+b_{32} c_5 (c_6 c_7+c_4 c_9)}{c_8},\nonumber\\
C_{12}&=&b_{31} c_2 c_3+b_{32} c_5 c_6+b_{33} c_8 c_9,\nonumber\\
C_{13}&=&\frac{2b_{33} c_7^2 c_9+b_{31} c_1(c_3 c_7+c_1 c_9)+b_{32} c_4 (c_6 c_7+c_4 c_9)}{c_9},\nonumber\\
C_{14}&=&\frac{2b_{33} c_8^2 c_9+b_{31} c_2(c_3 c_8+c_2 c_9)+b_{32} c_5 (c_6 c_8+c_5 c_9)}{c_9},\nonumber\\
C_{15}&=&b_{31} c_3^2+b_{32} c_6^2+b_{33} c_9^2,\nonumber\\
C_{16}&=&C_{6},\nonumber\\
C_{17}&=&\frac{2b_{33} c_7 c_8 c_9+b_{32} c_6(c_5 c_7+c_4 c_8)+b_{31} c_3 (c_2 c_7+c_1 c_8)}{c_9},\nonumber\\C_{18}&=&C_{12}.\end{eqnarray}

\end{document}